\documentclass[aps,pra,twocolumn,showpacs,superscriptaddress]{revtex4-2}  

\usepackage{physics}
\usepackage{graphicx}  
\usepackage{dcolumn}   
\usepackage{bm,amssymb,amsmath,dsfont}   
\usepackage[colorlinks=true,breaklinks=true,allcolors=blue]{hyperref}
\usepackage{url}
\usepackage{txfonts}
\usepackage{bbold}

\usepackage{graphics}
\usepackage{pstricks}
\usepackage{tikz}
\usepackage{color}
\usepackage{xcolor}

\usepackage{mathtools}

\usepackage{xr}
\renewcommand{\vec}[1]{\mathbf{#1}}
\newcommand{\eq}[1]{(\ref{eq:#1})}
\newcommand{\Eq}[1]{Eq.\,\eqref{eq:#1}}

\newcommand{\Fig}[1]{Fig.~\ref{fig:#1}}

\newcommand{\Sect}[1]{Sect.~\ref{sec:#1}}
\newcommand{\sect}[1]{\ref{sec:#1}}

\newcommand{\App}[1]{App.~\ref{app:#1}}

\DeclareMathOperator{\sgn}{sgn}
\newcommand{\ols}[1]{\mskip.5\thinmuskip\overline{\mskip-.5\thinmuskip {#1} \mskip-.5\thinmuskip}\mskip.5\thinmuskip} 

\newcommand{\df}[1]{\mathrm{d}^{d+1}\! #1 \,}

\newcommand{\tildeft}[1]{\tilde{#1}}
\renewcommand{\tildeft}[1]{#1}

\definecolor{applegreen}{rgb}{0.55, 0.71, 0.0}
\definecolor{byzantine}{rgb}{0.74, 0.2, 0.64}

\newcommand{\bae}[2]{#2} 

\makeatletter
\let\cat@comma@active\@empty
\makeatother

\begin{document}


\title{Anomalous scaling at non-thermal fixed points of the sine-Gordon model}
\author{Philipp Heinen}
\affiliation{Kirchhoff-Institut f\"ur Physik,
             Ruprecht-Karls-Universit\"at Heidelberg,
             Im~Neuenheimer~Feld~227,
             69120~Heidelberg, Germany}
\author{Aleksandr N. Mikheev}
\affiliation{Kirchhoff-Institut f\"ur Physik,
             Ruprecht-Karls-Universit\"at Heidelberg,
             Im~Neuenheimer~Feld~227,
             69120~Heidelberg, Germany}
\affiliation{Institut f\"{u}r Theoretische Physik,
		Universit\"{a}t Heidelberg, 
		Philosophenweg 16, 
		69120 Heidelberg, Germany}
\author{Thomas Gasenzer}
\affiliation{Kirchhoff-Institut f\"ur Physik,
             Ruprecht-Karls-Universit\"at Heidelberg,
             Im~Neuenheimer~Feld~227,
             69120~Heidelberg, Germany}
\affiliation{Institut f\"{u}r Theoretische Physik,
		Universit\"{a}t Heidelberg, 
		Philosophenweg 16, 
		69120 Heidelberg, Germany}

\date{\today}

\begin{abstract}
We extend the theory of non-thermal fixed points to the case of anomalously slow universal scaling dynamics according to the sine-Gordon model.
This entails the derivation of a kinetic equation for the momentum occupancy of the scalar field from a non-perturbative two-particle irreducible effective action, which re-sums a series of closed loop chains akin to a large-$N$ expansion at next-to-leading order.
The resulting kinetic equation is anal\bae{yse}{yze}d for possible scaling solutions in space and time that are character\bae{ise}{ize}d by a set of universal scaling exponents and encode self-similar transport to low momenta.
Assuming the momentum occupancy distribution to exhibit a scaling form we can determine the exponents by identifying the dominating contributions to the scattering integral and power counting.
If the field exhibits strong variations across many wells of the cosine potential, the scattering integral is dominated by the  scattering of many quasiparticles such that the momentum of each single participating mode is only weakly constrained.
Remarkably, in this case, in contrast to wave turbulent cascades, which correspond to local transport in momentum space, our results suggest that kinetic scattering here is dominated by rather non-local processes corresponding to a spatial containment in \emph{position} space.
The corresponding universal correlation functions in momentum and position space corroborate this conclusion.
Numerical simulations performed in accompanying work yield scaling properties close to the ones predicted here.
\end{abstract}

\maketitle
\setcounter{equation}{0}
\setcounter{table}{0}
\setcounter{page}{1}
\makeatletter

\section{\label{sec:Intro}Introduction}
Among the most challenging questions in classical and quantum field theory to date concerns the possible ways how interacting systems with many degrees of freedom evolve far out of equilibrium.
Non-thermal fixed points have been proposed to character\bae{ise}{ize} universal forms of far-from-equilibrium dynamics akin to renormal\bae{isation}{ization}-group fixed points that represent phase transitions in equilibrium \cite{Berges:2008wm,Berges:2008sr}.
During recent years, studies of universal phenomena far from equilibrium have intensified, in experiment
\cite{%
Navon2015a.Science.347.167N,
Navon2016a.Nature.539.72,
Johnstone2019a.Science.364.1267,
Eigen2018a.arXiv180509802E,
Prufer:2018hto,
Erne:2018gmz,
Navon2018a.doi:10.1126/science.aau6103,
Glidden:2020qmu,
GarciaOrozco2021a.PhysRevA.106.023314}
and theory
\cite{%
Berges:2008wm,
Berges:2008sr,
Scheppach:2009wu,
Schole:2012kt,
Nowak:2012gd,
Hofmann2014a,
Mathey2014a.PhysRevA.92.023635,
Maraga2015a.PhysRevE.92.042151,
Orioli:2015dxa,
Williamson2016a.PhysRevLett.116.025301,
Williamson2016a.PhysRevA.94.023608,
Bourges2016a.arXiv161108922B.PhysRevA.95.023616,
Chiocchetta:2016waa.PhysRevB.94.174301,
Karl2017b.NJP19.093014,
Schachner:2016frd,
Fujimoto2018a,
Walz:2017ffj.PhysRevD.97.116011,
Chantesana:2018qsb.PhysRevA.99.043620,
Mikheev:2018adp,
Schmied:2018upn.PhysRevLett.122.170404,
Mazeliauskas:2018yef,
Schmied:2018mte,
Fujimoto2018b.PhysRevLett.122.173001,
Williamson2019a.ScPP7.29,
Schmied:2019abm,
Gao2020a.PhysRevLett.124.040403,
Wheeler2021a.EPL135.30004,
Gresista:2021qqa,
RodriguezNieva2021a.arXiv210600023R,
Preis2023a.PhysRevLett.130.031602}, 
many of them in the field of ultracold gases.

These works have been performed in very different systems, while most of them have focused on the form and evolution of basic correlation functions such as distributions of particle momenta, structure factors, or phase coherence functions, which contain information about both, spatial and temporal variations in the system.
Universal properties such as power law exponents characterising these variations have been extracted with the aim to distinguish different universality classes.

Putting this recent activity into a broader perspective, studies of universal scaling far from equilibrium have been performed for more than a century. 
These include driven stationary systems such as turbulence in classical \cite{Frisch1995a,Kraichnan1967a} and quantum fluids \cite{Tsubota2008a, Vinen2006a}, wave turbulence \cite{Zakharov1992a,Nazarenko2011a}, as well as phase-ordering kinetics and coarsening phenomena following a quench into a phase with different possible order \cite{Bray1994a.AdvPhys.43.357,Puri2019a.KineticsOfPT,Cugliandolo2014arXiv1412.0855C}.
Turbulence phenomena are typically anal\bae{yse}{yze}d with respect to their scaling properties on the grounds of hydrodynamic and kinetic equations. 
In contrast, the standard classification of phase-ordering kinetics builds on that of dynamical critical scaling in the linear response of systems close to equilibrium \cite{Hohenberg1977a,Bray2003a.rsta.2002.1164}, and of non-linear critical relaxation \cite{Racz1975a.PLA.53.433,Fisher1976a.PhysRevB.13.5039,Bausch1976b,Bausch1979a}.

Phase-ordering kinetics and coarsening, though, as many transient turbulent motions, typically result from a quench far out of equilibrium and exhibit self-similar evolution.
Take, e.g., spin domains formed in a shock-cooled magnet, which on long time scales grow according to a power law in time, $\ell_{\mathrm{d}}(t)\sim t^{\,\beta}$, with a universal scaling exponent $\beta$.
As in stationary situations like fully developed turbulence, the momentanous spatial distribution of the order parameter field is often character\bae{ise}{ize}d by a so-called Porod tail in momentum space, $f(t,p)\sim p^{-\kappa}$, with universal exponent $\kappa$. 
Universality, here, means that the spatio-temporal form of this distribution as well as of more general statistical correlations is independent of the particular physical real\bae{isation}{ization} but rather reflects characteristic symmetries and related conservation laws. 

Since far from equilibrium, a system is subject to less constraints and thus can show distinctly different scaling phenomena, a unified description of such phenomena is desirable, but lacking so far for most practically relevant cases of coarsening \cite{Cugliandolo2014arXiv1412.0855C}.
Non-thermal fixed points offer such a unified description provided it is possible to apply the relevant techniques to the strongly correlated excitations governing such phenomena.

Here we follow the strategy to make use of techniques, that have been developed for weak wave turbulence \cite{Zakharov1992a,Nazarenko2011a} and which have been extended to characterise non-thermal fixed points and capture strong wave turbulence \cite{Berges:2008wm,Berges:2008sr,Scheppach:2009wu,Orioli:2015dxa,Chantesana:2018qsb.PhysRevA.99.043620}. 
We further extend these analytical methods for describing scaling phenomena driven by strongly non-linear interactions, of non-polynomial form. 
As an example, we consider a non-linear wave equation with sinusoidal interactions governing a scalar field, which constitutes the so-called sine-Gordon model.
The sine-Gordon model is highly relevant in many different contexts, e.g., through its mapping to the Thirring model \cite{Coleman1975a.PRD11.2088} and its soliton and kink solutions  \cite{Gogolin2004a.bosonization,Giamarchi2003a,Cuevas2014a.sineGordon}, its mapping to a Coulomb gas \cite{Edwards1962a.JMP.3.778,Chui1975a.PhysRevLett.35.315,Samuel1978a.PhysRevD.18.1916,Amit1980a.JPA13.585,Nandori2022a.NPB975.115681}, which is useful in describing the Berezinskii-Kosterlitz-Thouless transition in two-dimensional Bose systems \cite{Minnhagen1987a.RevModPhys.59.1001,Berezinskii1971JETP...32..493B,Berezinskii1972JETP...34..610B,Kosterlitz1973a,Kosterlitz2020a.JLTP}, also in one spatial dimension at finite temperature \cite{Jose1977a.PhysRevB.16.1217} and out of equilibrium \cite{Barmettler2010a,Nogueira2005a.PhysRevB.72.014541,Nogueira2006a.PhysRevB.73.104515}, for structure formation and growth in the universe \cite{Turok1991a.PhysScr,Greene:1998pb,Berges:2014xea,Berges:2019dgr,Lentz2019a.MNRAS485.1809}, quark confinement \cite{Polyakov1977a.NPB120.429.quark} as well as for false vacuum decay \cite{Hawking1982a.PhysRevD.26.2681,Braden2015a.JCAP03.007}.
Numerical simulations of coarsening within this model, as presented in \cite{Heinen:2022rew}, exhibit anomalously slow scaling in space and time, corroborating the analytical results reported here.

This opens a perspective on analytically capturing coarsening phenomena, which so far have been beyond the reach of ab-initio methods as used here.
We expect our methods to be applicable to more general types of models, which could lead to a classification of the universal properties that character\bae{ise}{ize} domain coarsening and phase-ordering kinetics.
We derive kinetic equations on the grounds of functional quantum field theory and anal\bae{yse}{yze} their scaling solutions within a classical statistical approximation that is valid for large occupancies, when effects arising from fluctuations at the single-particle level can be neglected.
The chosen approach allows us taking into account the self-interactions of the field in a non-perturbative manner, which means that the kinetic equation remains applicable to arbitrarily large occupancies of long-wavelength modes that character\bae{ise}{ize} ordering phenomena.
Distinctly different from stationary (wave) turbulent transport, we find the scattering processes that drive the evolution at the non-thermal fixed point to be strongly non-local in momentum space. 
As a result, the corresponding correlations are rather local\bae{ise}{ize}d in position space, which appears to be a prerequisite for the type of pattern formation seen in the simulations of the system \cite{Heinen:2022rew}.

Our paper is organi\bae{s}{z}ed as follows:
In \Sect{kineticSG}, we introduce the kinetic description of the sine-Gordon model on the basis of functional field theoretic techniques, derive dynamic equations in a non-perturbative approximation scheme, and reduce these to a wave-Boltzmann-type equation for the momentum distribution of the system.
In \Sect{scalingSG}, we perform a scaling analysis of the resulting kinetic equation in space and time and derive scaling exponents at a new type of infrared non-thermal fixed point and discuss its properties in view of coarsening of non-homogeneous field patterns in position space.
A summary an outlook is given in \Sect{Conclusions}.

\section{\label{sec:kineticSG}Kinetic theory of the sine-Gordon model}
In this first chapter, we derive the kinetic equation for the momentum distribution function $f(t,\mathbf{p})$ of the sine-Gordon model within a non-perturbative approximation.
The procedure developed here extends the standard procedures \cite{Berges:2015kfa,Berges:2004yj,Lindner:2005kv,Lindner:2007am} by including the non-polynomial, cosine interaction potential of the model.
Our further goal is to determine possible self-similar evolution of correlation functions of the sine-Gordon model, in which kinetic equations describing transport in momentum space are anal\bae{yse}{yze}d for possible scaling solutions in space and time, close to a non-thermal fixed point, see, e.g., Refs.~\cite{Berges:2008wm,Scheppach:2009wu,Berges:2010ez,Orioli:2015dxa,Berges:2015kfa,Chantesana:2018qsb.PhysRevA.99.043620}. 

In \Sect{sGmodel}, we introduce the model, then start from the Kadanoff-Baym dynamic equations for two-point correlators (\Sect{NoneqEvolEqs}), from which we derive the wave-Boltzmann type equation within a first-order gradient expansion (\Sect{KinEq}).
The scattering integral of the resulting kinetic equation is non-perturbative and can be obtained from a two-particle irreducible (2PI) effective action or $\Phi$-functional (\Sect{SE}).
For the description of an infrared non-thermal fixed point, we will use, as successfully explored for standard $\lambda\phi^{4}$ field models \cite{Berges:2008wm,Scheppach:2009wu,Berges:2010ez,Orioli:2015dxa,Berges:2015kfa,Chantesana:2018qsb.PhysRevA.99.043620}, a non-perturbative approximation of the self-energies entering the dynamic equations.
This includes, beyond the standard approach, the re-summation of propagator lines for the sine-Gordon interactions, which involve arbitrary orders $n$ in the field interactions $\varphi^{n}$ (\Sect{ResumLinesSG}), and an $s$-channel re-summation of a set of 2PI loop-chain diagrams involving the corresponding high-order vertices (Sects.~\sect{Resum2PIEASG}--\sect{KBEq}).
This eventually leads to the wave-Boltzmann scattering integral and $T$-matrix involving arbitrary powers of the distribution function $f$ (\Sect{ScattInt}).
Specifically, in the limit of large occupancies, as they are anticipated, in the scaling limit, for low-momentum modes, the $T$-matrix becomes independent of the bare coupling constant of the model but rather depends in a transcendental manner on the distribution function only.

\subsection{Sine-Gordon Model}
\label{sec:sGmodel}
The `sine-Gordon' equation, which is a non-linear Klein-Gordon equation with a sine-function non-linearity,
\begin{equation}
	\label{eq:sGE}
	\Box\varphi + m^{2} \sin \varphi = 0\,, 	
\end{equation}
can be derived as an Euler-Lagrange equation from the Lagrangian density
\begin{equation}
	\label{eq:SG_lagrangian}
	\mathcal{L}_{\mathrm{sG}} 
	=
	\frac{1}{2\eta}\partial_{\mu} \varphi\,\partial^{\,\mu} \varphi + \lambda \left(\cos\varphi - 1\right), 
\end{equation}
which is typically written in terms of the real coupling parameters $\lambda$ and $\eta$, with $\eta\lambda=m^{2}$.
The parameter $\eta$ has the mass dimension $[\eta]=1-d$ in $d$ spatial dimensions, which is chosen such that the real scalar field $\varphi$ is dimensionless.
From the above Lagrangian one can readily derive the inverse of the free propagator,
\begin{equation}
	\label{eq:G_0}
	\mathrm{i} G^{-1}_0(x,y) 
	\equiv \left.\frac{\delta^2 S[\varphi]}{\delta\varphi(x)\delta\varphi(y)}\right\rvert_{\phi_0} 
	= -\eta^{-1}\left[\Box_x + m^2 \right] \delta_{\mathcal{C}}(x - y)\,.	
\end{equation}

The coupling $\lambda$, with $[\lambda]=1+d$,  sets the strength of the interactions.
Writing the cosine potential as its Taylor series, it follows that the coupling constants of all vertices, from second to arbitrarily high order, are fixed by a single parameter, $m^{2}$.  
Note that one may also rescale the dimensionless field $\varphi$ as $\varphi=\phi\sqrt{\eta}$, such that $\phi$ carries the same dimension as in the Klein-Gordon model. 
The sine-Gordon Lagrangian then reads $\mathcal{L}=(\partial\phi)^{2}/2 + \eta^{-1}m^{2}[\mathrm{cos}(\sqrt\eta\,\phi)-1]$, which shows that $\eta$ controls the relative weight of the higher-order vertices, e.g., of the standard $\lambda_{4}\phi^{4}/4!$ vertex with coupling constant $\lambda_{4}=-\eta m^{2}=-\eta^{2}\lambda$, as compared with $m^{2}\phi^{2}/2$.

The sine-Gordon model plays an important role in equilibrium, in providing an effective description of topological defects such as vortices \cite{Edwards1962a.JMP.3.778,Minnhagen1987a.RevModPhys.59.1001,Nagaosa2013.QFTinCMPhysics,Kleinert1995cond.mat.9503030K,Antonov1999a.IJMPA27.4347.abelian,Nogueira2006a.PhysRevB.73.104515,Defenu2021topological}.
For the non-equilibrium case, such a connection has not been established so far (cf., however, the discussion in \Sect{Conclusions}). Nonetheless, the sine-Gordon model finds application also in numerous out-of-equilibrium settings, such as in the dynamics of axions \cite{Berges:2014xea} or two coupled Bose-Einstein condensates \cite{Bouchoule2005a.EPJD35.147.modulational}.
The dynamics of the sine-Gordon model after quenches has been studied in, e.g., \cite{Sabio2010a.NJP12.055008,DallaTorre2013.PhysRevLett.110.090404,Horvath2019a.PhysRevA.100.013613}. 

We finally remark that we are here mainly interested in the universal far-from-equilibrium dynamics of the sine-Gordon model in the form of transport towards the infrared.
Hence, we consider the sine-Gordon model as a low-energy effective field theory.
This does not exclude, however, that issues of renormali\bae{s}{z}ability \cite{Nagy2007a.PLB647.152,Faber2003a.JPA36.7839} could give rise to corrections to the approximate results obtained in the following, which are, however, beyond the scope of the present work.

\subsection{Nonequilibrium evolution equations for two-point correlation functions}
\label{sec:NoneqEvolEqs}
The sine-Gordon model is invariant under translations of the field $\varphi(x)\to\varphi(x)+2\pi n$ by integer multiples of $2\pi$, such that also the field expectation value $\langle\varphi\rangle$ is defined only up to these integer multiples.
In the following, we assume that, in the statistical average over many configurations, the field expectation value is and stays zero at every spatial position in the system.
This assumption is justified as long as there is no explicit symmetry breaking nor any non-vanishing mean field in the initial condition.
And it does not exclude spontaneous or any other dynamical symmetry breaking as long as the average is taken over all different possible solutions according to the right probability distribution, cf., e.g.~\cite{Orioli:2015dxa,Schachner:2016frd}.
This means that topologically non-trivial objects such as kinks can be captured by the description in terms of two-point and higher-order correlation functions only.
Notwithstanding this, one needs to keep in mind that kinks, e.g., are not necessarily accounted for within the approximations made in the following.
As we will point out at the end, however, numerical results are remarkably close to the scaling exponents predicted here. 

In consequence, we will restrict our analysis to two-point correlators the evolution of which is governed by the Kadanoff-Baym equations of motion  \cite{Berges:2015kfa},
\begin{subequations}
\label{eq:KB}
\begin{align}
	\label{eq:KB_F}	
	\eta^{-1}\left[\Box_x + M^2(x)\right] F(x,y) 
	=&
	-\int_{t_0}^{x^0} \dd{z} \Sigma^{\,\rho}(x,z) F(z,y)
	\\ \nonumber
	&+\ \int_{t_0}^{y^0} \dd{z} \Sigma^F(x,z)\rho(z,y)\,,\\
	\label{eq:KB_rho}
	\eta^{-1}\left[\Box_x + M^2(x)\right] \rho(x,y)
	&=
	-\int_{y^0}^{x^0}\dd{z} \Sigma^{\,\rho}(x,z) \rho(z,y)\,,
\end{align}	
\end{subequations}
for the statistical, $F(x,y)=\left\langle\left\lbrace\hat{\varphi}(x),\hat{\varphi}(y)\right\rbrace\right\rangle/2$, and spectral functions, $\rho=\mathrm{i} \left\langle\left[\hat{\varphi}(x),\hat{\varphi}(y)\right]\right\rangle$, where the renormal\bae{ise}{ize}d mass $M$ is defined in \Eq{RenMass} below and we chose the notation $\int_{t_1}^{t_2}\dd{z} \equiv \int_{t_1}^{t_2}\dd{z^0}\int\dd[d]{\mathbf{z}}$. 
$F$ and $\rho$ derive from the time-ordered connected Green's  function
\begin{align}
\label{eq:G_decomp}
	G(x,y) 
	= \langle \mathcal{T}_{\mathcal{C}}\hat{\varphi}(x)\hat{\varphi}(y)\rangle
	= F(x,y) - \mbox{$\frac{\mathrm{i}}{2}$} \rho(x,y) \sgn_{\mathcal{C}}(x^0 - y^0)
	\,,
\end{align}
where $\mathcal{C}$ denotes the Schwinger-Keldysh closed-time path.
The self-energy $\Sigma$, which contains all information about the structure of correlations induced by the non-linear interactions and will be discussed further in the next-but-one section, has been decomposed into a local and a non-local part, 
\begin{equation}
\label{eq:self_energy_loc_nonloc}
	\Sigma(x,y) \equiv -\mathrm{i} \Sigma^{(0)}(x)\,\delta_{\mathcal{C}}(x-y) + \overline{\Sigma}(x,y)\,,
\end{equation}
where the local part, together with the bare mass, defines the dynamic mass squared 
\begin{equation}
\label{eq:RenMass}	
	M^2(x) \equiv m^2 + \Sigma^{(0)}(x)
\end{equation}
on the left-hand side of Eqs.~\eq{KB}. 
The non-local part has been decomposed, in analogy to $G$,
into a statistical and a spectral part, 
\begin{equation}
\label{eq:self_energy_decomp}	
	\overline{\Sigma}(x,y) 
	\equiv \Sigma^F(x,y) - ({\mathrm{i}}/{2}) \Sigma^{\,\rho}(x,y) \sgn_{\mathcal{C}}(x^0 - y^0)
	\,.
\end{equation}
%

\subsection{Kinetic equation}
\label{sec:KinEq}
For the later scaling analysis, we consider a uniform, translationally invariant system and go over to a description in momentum space.
The scaling limit is reached at late times and low wave numbers, implying an algebraic slow-down of the evolution of $G(x,y)$ along the central-time direction $\sim x^0 + y^0$.
This suggests an approximate description that takes into account only low orders in an expansion in both, temporal and spatial derivatives with respect to the central coordinates $X\sim x+y$. 
Along the standard line of arguments, a transport equation that governs the late-time dynamics is obtained by applying such a gradient expansion to the dynamic equations \eq{KB}, with the initial time sent to $t_0 \to -\infty$ \cite{Berges:2015kfa,Branschadel:2008sk,Berges:2004yj}. 
At leading order of the gradient expansion, one obtains
\begin{subequations}
\begin{align}
	\label{eq:F_tr_eq}	
	\frac{2\mathrm{i}p^{\,\mu}}{\eta} \pdv{F(X,p)}{{X^{\mu}}} 
	&=
	\tildeft{\Sigma}^{\,\rho}(X,p) F(X,p) - \Sigma^F(X,p) \tildeft{\rho}(X,p)
	\,,\\
	\label{eq:rho_tr_eq}
	\frac{2\mathrm{i}p^{\,\mu}}{\eta} \pdv{\tildeft{\rho}(X,p)}{X^{\mu}}
	&=
	0
\end{align}
\end{subequations}
governing the center-time evolution of the Fourier transforms in the relative coordinate,
\begin{subequations}
\label{eq:FrhoXp}
\begin{align}
	\label{eq:FXp}
	F(X,p) 
	&=
	\int\dd[d+1]{s} \mathrm{e}^{\mathrm{i} p \cdot s} F\left(X + \frac{s}{2},X - \frac{s}{2}\right)
	\,,\\
	\label{eq:rhoXp}
	\tildeft{\rho}(X,p)
	&=
	\int\dd[d+1]{s} \mathrm{e}^{\mathrm{i} p \cdot s} \rho\left(X + \frac{s}{2},X - \frac{s}{2}\right),
\end{align}
\end{subequations}
with $p$, $s$, and $X$ $\in \mathbb{R}^{1+d}$ such that $p\cdot s = p^{0} s^0 - \mathbf{p}\cdot \mathbf{s}$. 
The self-energies $\Sigma^F(X,p)$ and $\tildeft{\Sigma}^{\,\rho}(X,p)$ are defined analogously. 

To describe transport by means of kinetic theory, it is convenient to introduce the concept of quasiparticles. 
In the simplest quasiparticle approximation, one takes into account only the linear part of the Kadanoff-Baym equation \eq{KB_rho} for the spectral function, the solution of which is the free spectral function of the sine-Gordon model,
\begin{align}
	\label{eq:rho_free}
	\tildeft{\rho}^{(0)}(p)
	&=
	2\pi\mathrm{i}\eta\sgn\left(p^0\right)\,\delta\left(\left[p^0\right]^2 - \omega_{\mathbf{p}}^2\right)
	\nonumber\\
	&= 
	\frac{\mathrm{i}\pi\eta}{\omega_{\mathbf{p}}}\,\left[\delta\left(p^0 - \omega_{\mathbf{p}}\right)-\delta\left(p^0 + \omega_{\mathbf{p}}\right)\right]
	\,, 
\end{align}
with dispersion depending on the mass squared \eq{RenMass},
\begin{equation}
	\label{eq:omegap_free}
	\omega_{\mathbf{p}}=\omega^{(0)}_{\mathbf{p}}
	= (\mathbf{p}^{2}+M^{2})^{1/2}
	\,.
\end{equation}
This spectral function is then used to solve the fully non-linear equation \eq{F_tr_eq} for the statistical function $F$.
Without loss of generality, one may express $F$ in terms of $\rho$ by
\begin{equation}
	\label{eq:FDR}	
	F(X,p) = -\mathrm{i}\left[f(X,p) + \frac{1}{2}\right] \tildeft{\rho}(X,p)\,,
\end{equation}
and note that, for a real scalar theory, one has
\begin{equation}
	\label{eq:FrhoSymmetries}
	F(X,-p) = F(X,p)\,, \quad \tildeft{\rho}(X,-p) = -\tildeft{\rho}(X,p)\,,
\end{equation}
which implies that $f(X,-p) = -\left[f(X,p) + 1\right]$.
For our spatially homogeneous system, the spectral function is constant in central time $t\equiv X^{0}$, cf.~\Eq{rho_tr_eq}, while the statistical function $F$ is not, and both functions are independent of $\mathbf{X}$.
Integrating \Eq{F_tr_eq} over the frequency $p^{0}$ and using the parametr\bae{isation}{ization} \eq{FDR} we obtain the wave-Boltzmann-type kinetic equation
\begin{equation}
	\label{eq:pre_kinetic}
	\partial_t f(t,\mathbf{p}) = C[f](t,\mathbf{p})\,,
\end{equation}
for the quasiparticle distribution function
\begin{equation}
	\label{eq:f_def}
	f(t,\mathbf{p}) 
	\equiv -\mathrm{i}\int_0^{\infty} \frac{\dd{p^0}}{2\pi} 2 p^0 \eta^{-1} \tildeft{\rho}(p) f(t,p)\,.
\end{equation}
with the scattering integral
\begin{equation}
	\label{eq:Cf}
	C[f](t,\mathbf{p})
	=
	-\mathrm{i}\int_0^{\infty} \frac{\dd{p^0}}{2\pi} \left[\tildeft{\Sigma}^{\,\rho}(t,p) F(t,p) - \Sigma^F(t,p) \tildeft{\rho}(t,p)\right]
	\,.
\end{equation}
For the free spectral function \eq{omegap_free} the integral \eq{f_def} implies, with \eq{rho_free}, that the function $f(t,\mathbf{p})$ is evaluated on-energy-shell, $f(t,\mathbf{p}) = f(t,p^0=\omega_{\mathbf{p}},\mathbf{p})$.
This function describes the distribution of stable quasiparticles over the momenta $\mathbf{p}$, with the momentum-dependent energy $\omega_{\mathbf{p}}$ being determined by the quadratic part of the Lagrangian.

\subsection{Approximation scheme for the self-energies}
\label{sec:SE}
In order to close the set of equations \eqref{eq:KB}, we  specify the self-energy $\Sigma$ self-consistently within an expansion of the effective action in terms of two-particle irreducible (2PI) diagrams \cite{Berges:2015kfa}.
The relevant part of the effective action is defined as 
\begin{equation}
\label{eq:gamma2}	
	\Gamma_2[G] 
	= -\mathrm{i}\ln\left\langle \mathrm{e}^{\mathrm{i}S_{\mathrm{int}}[\varphi]}\right\rangle_{2\mathrm{PI}}
	\,,
\end{equation}
where
\begin{equation}
\label{eq:Sint}	
	S_{\mathrm{int}}[\varphi]
	= -\int_{\mathcal{C}}\mathrm{d}z\,\mathcal{V}(\varphi)
\end{equation}
represents the interaction part of the action, in our case in terms of the field potential
$\mathcal{V}(\varphi)=\lambda(\cos\varphi-1)$. 
The double Legendre transform giving the 2PI effective action or Luttinger-Ward $\Phi$-functional \cite{Luttinger1960a,Baym1962a,Cornwall1974a} as a functional of  $G$ can be formulated as the logarithm of a (time-ordered) expectation value of the exponentiated $S_{\mathrm{int}}$, where only two-particle irreducible contributions (with full propagators and bare vertices) are kept.
The beyond-1-loop part $\Gamma_{2}$ defines the self-energy,
\begin{equation}
\label{eq:self_energy_def}	
	\Sigma(x,y;G) \equiv 2\mathrm{i} \frac{\delta\Gamma_2[G]}{\delta G(x,y)}
	\,.
\end{equation}

The spatial scaling properties at an infrared non-thermal fixed point \cite{Berges:2008wm,Scheppach:2009wu} as well as the corresponding dynamical scaling evolution \cite{Orioli:2015dxa,Chantesana:2018qsb.PhysRevA.99.043620}, have been predicted, for scalar field theories,  on the grounds of a large-$N$ expansion of the 2PI effective action, which corresponds to an $s$-channel bubble re-summation of $\Gamma_{2}$ akin to the random-phase and GW approximations \cite{Hedin1965a.PhysRev.139.A796,
Aryasetiawan1998a.RepProgPhys61.237,
Stan2009a.JChemPhys130.114105.levels,
Ren2012aJMatS47.7447}.

A system described by a relativistic $\mathcal{O}(N)$-symmetric scalar model with quartic interactions $\sim\lambda(\phi_{a}\phi_{a})^{2}$ is naturally dominated by the fluctuations of the $N-1$ Goldstone degrees of freedom while fluctuations of the single radial, ``Higgs'' mode are energetically suppressed by the quartic interaction term.
A similar situation prevails in the case of a non-relativistic $\mathcal{U}(N)$-symmetric Schr\"odinger model with a quartic non-linearity  $\sim g(\psi^{\dagger}_{a}\psi_{a})^{2}$.
Its dynamics is dominated by the $N-1$ independent Goldstone modes with free, i.e., quadratic dispersion, which are character\bae{ise}{ize}d by equal-magnitude excitations of the relative phases and amplitudes between the different components. 
At the same time, the long-wave-length excitations of the total density $\rho=\psi^{\dagger}_{a}\psi_{a}$ are energetically suppressed.
They form another Goldstone excitation with a sound-like dispersion, dominated by strong phase variations, while there is no massive Higgs mode.

For both types of systems, in the large-$N$ limit, the dynamics of the strongly occupied low-momentum modes, building up as a result of the transport of particles towards the infrared when the system is close to a non-thermal fixed point, are thus anticipated to be dominated by the interactions of the $N-1$ Goldstone excitations with free dispersion.
The above mentioned large-$N$ expansions at next-to-leading order capture this dynamics while they take into account the suppressed massive Higgs or sound-like total-density degrees of freedom in a static manner only.
This picture has been confirmed by an analysis within a low-energy effective theory approach, in which, after integrating out the subdominant radial density fluctuations, the same universal scaling behavi\bae{our}{or} emerged as in the fundamental field theory  \cite{Mikheev:2018adp}. 

Here, we choose a similar approach 
and consider an $s$-channel type re-summation scheme, while still applying it to a model with a single scalar field only.
In the following, we work out this re-summation scheme and show that a closed-form expression results for the 2PI effective action and thus for the self-energies and the ensuing dynamic and kinetic equations.

\subsection{Re-summed propagator lines for the sine-Gordon model}
\label{sec:ResumLinesSG}
For deducing the above mentioned closed-form expression of an approximated 2PI effective action, we need to discuss in some more detail possibilities for re-summing 2PI diagrams for the sine-Gordon model. 
It is known from the discussion of higher-order terms in the BKT flow equations how to re-sum the infinite number of vertices from the cosine potential at a fixed order in the coupling  $\lambda$ \cite{Minnhagen1978a.PhysRevB.18.1356,Samuel1978a.PhysRevD.18.1916,Amit1980a.JPA13.585}. 
On this basis, we will introduce, in \Sect{Resum2PIEASG}, a re-summation of a class of diagrams to all orders in $\lambda$.

\begin{figure}[tb]
	\centering
	\includegraphics[width=0.425\columnwidth]{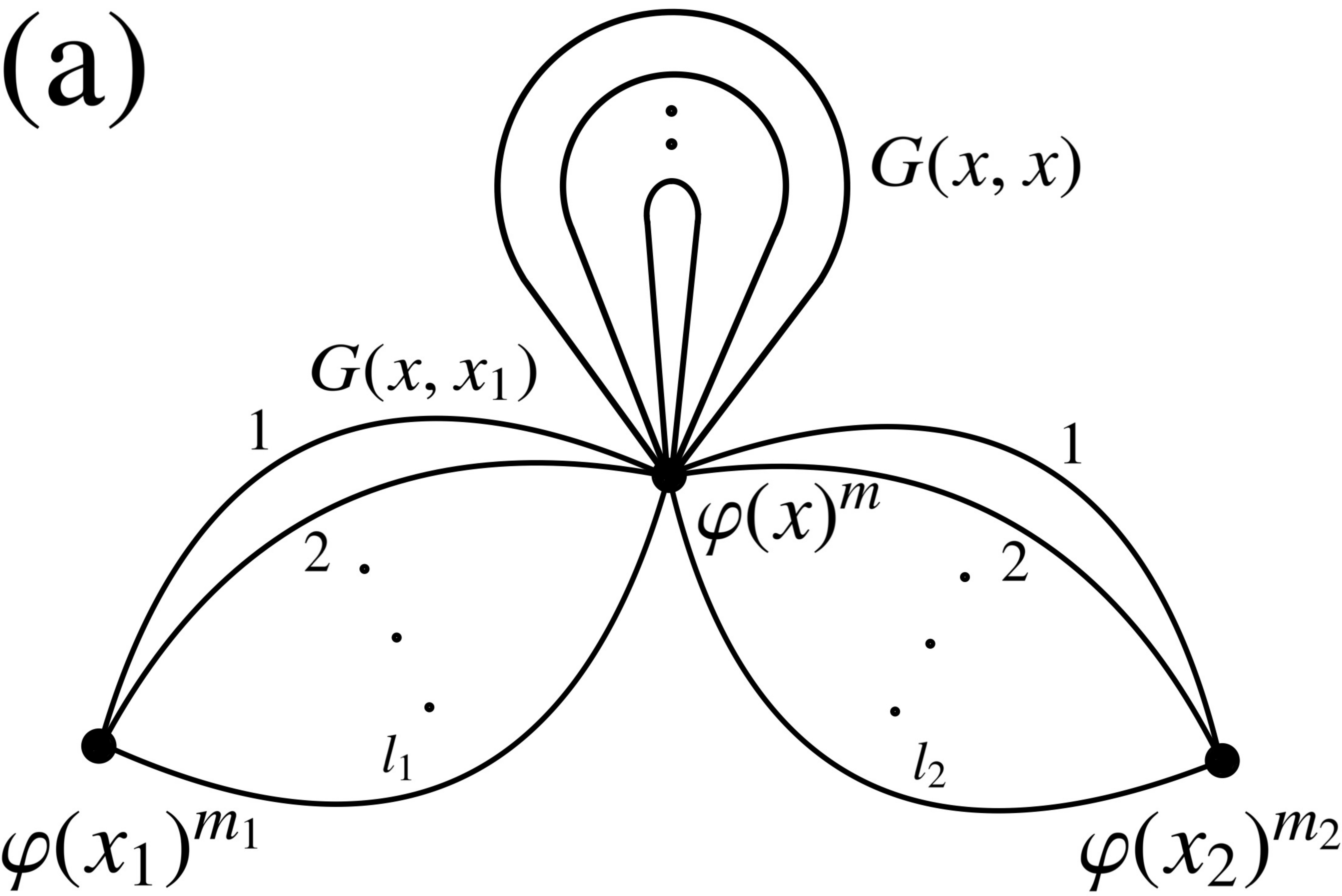}
	\includegraphics[width=0.525\columnwidth]{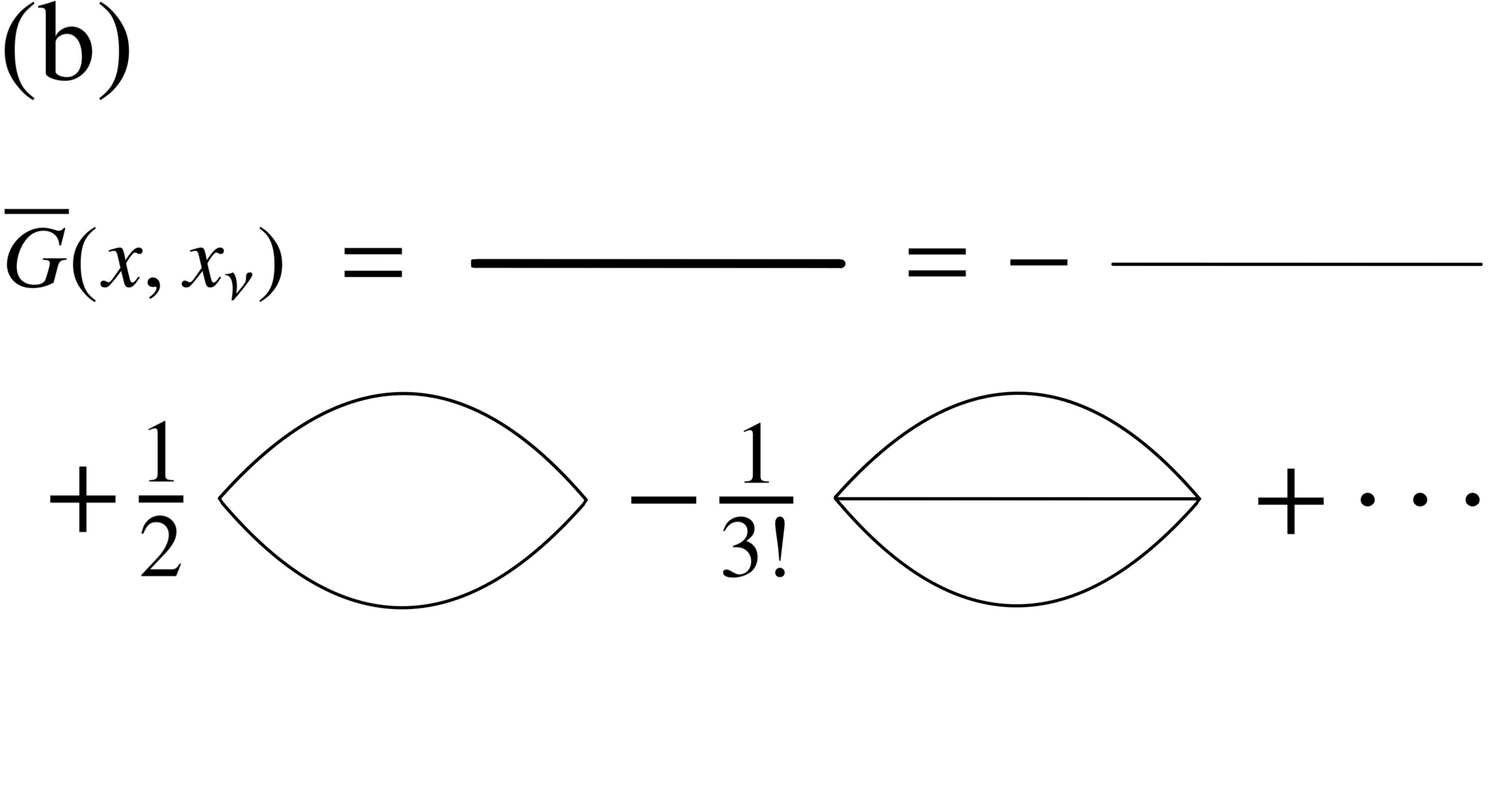}
	\caption{(a) A single $m$-vertex at space-time point $x$ is connected,  by $l_{1}$ of its legs, via full propagators $G(x,x_{1})$, to a neighb\bae{our}{or}ing vertex $\sim\varphi(x_{1})^{m_{1}}$, which we label by $1$, by $l_{2}$ legs to a different vertex $2$, and so on to $l_{L}$ legs linked to vertex $L$.
	In addition to this, $l_{0}=2l$ legs of each vertex are connected in a pair-wise manner to each other by local propagators $G(x,x)$.
	The multiplicities sum up to $\sum_{\nu=0}^{L}l_{\nu}=m$.
	(b) Re-summation of the diagrammatic series over powers $[G(x,x_{\nu})]^{l_{\nu}}$, within the otherwise identical 2PI diagrams.
	Such a re-summed link $\overline{G}_\mathrm{nl}(x,x_{\nu})$ is denoted by a thick line.
	The combinatorial factors indicate the non-local line \eq{evenoddResummedLinesSG}, which consists of even and odd contributions of opposite sign.
	There are no vertex couplings included in this re-summation.
	}
	\label{fig:LinkResummation}
\end{figure}

To start with, we Taylor expand the cosine potential,
\begin{equation}
\label{eq:Vexpansion}	
	\mathcal{V}(\varphi) 
	= \lambda(\cos\varphi-1) 
	= \sum_{m} \left[\frac{\mathcal{V}^{(m)}(\phi)}{m!}\right]\varphi^{m}
	\equiv \sum_{m} {v}_{m}\varphi^{m}\,,
\end{equation}
with $v_{m}=(-1)^{m/2}\lambda/m!$ for $m=2n$, $n\in\mathbb{N}$, while $v_{m}=0$ for all other integers $m$.
Now let us consider a single $m$-vertex at space-time point $x$ and assume that $l_{1}$ of its legs are connected, via full propagators $G(x,x_{1})$, to a neighb\bae{our}{or}ing vertex $\sim\varphi(x_{1})^{m_{1}}$, which we label by $1$, $l_{2}$ legs to a different vertex $2$, and so on to $l_{L}$ legs linked to vertex $L$.
In addition to this, $l_{0}=2l$ legs be connected in a pair-wise manner to each other by local propagators $G(x,x)$, see \Fig{LinkResummation}(a).
The multiplicities sum up to $\sum_{\nu=0}^{L}l_{\nu}=m$.

%
%
Since all the propagators $G(x,x_{\nu})$, $\nu=0,\ldots,L$, $x_{0}\equiv x$, within a $l_{\nu}$-multiple link are identical, our aim is to be able to re-sum the series over powers $[G(x,x_{\nu})]^{l_{\nu}}$, within the otherwise identical 2PI diagrams.
We denote such a re-summed link $\overline{G}(x,x_{\nu})$ by a thick line, see \Fig{LinkResummation}(b).
Such re-summations are in principle possible since each link $\overline{G}(x,x_{\nu})$ can be associated with a specific vertex, say that one at $x$. 
Moreover, the summation over the powers $[G(x,x_{\nu})]^{l_{\nu}}$ of lines can be included in the summation over the number of legs $m$.
To show this, let us first consider the combinatorics:
There are 
$m!/[(2l)!(l_{1}+\ldots+l_{L})!]$ indistinguishable ways to pick out the $2l$ from the $m$ legs that are locally pairwise connected by $G(x,x)$.
Then, there are $(l_{1}+\dots+l_{L})!/(l_{1}!\dots l_{L}!)$ ways how to connect the remaining lines via $[G(x,x_{\nu})]^{l_{\nu}}$ to the neighb\bae{our}{or}ing vertices $\nu=1,\ldots,L$. 
Finally, there are $(2l)!/(2^{l}l!)$ and $l_{\nu}!$ ways to permute the local and the non-local lines connecting to vertex $\nu$, respectively.
For each vertex, the Taylor coefficient is thus replaced by 
\begin{align}
\label{eq:CombinatoricsFactor}
	\frac{\mathcal{V}^{(m)}(\phi)}{m!}	
	&\frac{m!}{(2l)!(l_{1}+\ldots+l_{L})!}
	\frac{(l_{1}+\dots+l_{L})!}{l_{1}!\dots l_{L}!}
	\frac{(2l)!}{2^{l}l!}
	\prod_{\nu=1}^{L}\sqrt{l_{\nu}!}
	\nonumber\\
	&=\frac{\mathcal{V}^{(2l+l_{1}+\ldots+l_{L})}}{l!2^{l}(l_{1}!\dots l_{L}!)^{1/2}}
	\,.
\end{align}
Here, the multiplicity of the non-local connections between vertices $0$ and $\nu$ has been included as the square root of $l_{\nu}$ to avoid squared counting. 
Hence, to be able to sum over the multiplicities $l,l_{1},\ldots,l_{L}$, the numerator $\mathcal{V}^{(2l+l_{1}+\ldots+l_{L})}(\phi)$ needs to factor\bae{ise}{ize}, up to some constant $\lambda$, into components depending in a specific fashion on the $l_{\nu}$,
\begin{equation}
\label{eq:VmFactorisation}	
	\mathcal{V}^{(2l+l_{1}+\ldots+l_{L})}(\phi)
	=\lambda\, g_\mathrm{l}(l)\prod_{\nu=1}^{L}\sqrt{g_\mathrm{nl}(l_{\nu})}
\,,
\end{equation}
where the (squared) factors $g_\mathrm{l}$ and $g_\mathrm{nl}$ form the coefficients of the re-summed local and non-local lines, respectively,
\begin{subequations}
\begin{align}
\label{eq:ResummedLines}
	\overline G_\mathrm{l}(x,x) 
	&= \sum_{l=0}^{\infty}\frac{g_\mathrm{l}(l)}{l!}\left(\frac{G(x,x)}{2}\right)^{l}
	\,,
	\\
	\overline G_\mathrm{nl}(x,x_{\nu}) 
	&= \sum_{l_{\nu}=0}^{\infty}\frac{g_\mathrm{nl}(l_{\nu})}{l_{\nu}!}\left[G(x,x_{\nu})\right]^{l_{\nu}}
	\,.
\end{align}
\end{subequations}
Note that the factors depending on the indices $l_{\nu>0}$ need to be squared because an identical such factor arises from each of the linked vertices at $x$ and $x_{\nu}$.
Since $\mathcal{V}^{(m)}(\phi)=(-1)^{m/2}\lambda$ for $m=2n$, $n\in\mathbb{N}$, and $\mathcal{V}^{(m)}(\phi)=0$ for all other $m$, the factor\bae{isation}{ization} condition \eq{VmFactorisation} is fulfilled by the sine-Gordon model, and both, $g_\mathrm{l}(l)=(-1)^{l}$ and $g_\mathrm{nl}(l_{\nu})=(-1)^{l_{\nu}}$, result as the same pure, real phase, giving
\begin{subequations}
\begin{align}
	\label{eq:ResummedLinesSG}
	\overline G_\mathrm{l}(x,x) 
	&= \exp\left[-G(x,x)/2\right]
	\,,
	\\
	\overline G_\mathrm{nl}(x,x_{\nu}) 
	&= \exp\left[-G(x,x_{\nu})\right]
	\,.
\end{align}
\end{subequations}

Note however, that these results for the sine-Gordon model require that, in a given diagram, an even number of propagator lines $G(x,y)$ is attached to each vertex $\sim\varphi^{m}$.
Hence, one needs to distinguish the partial sums over even and odd powers of $G(x,x_{\nu})$,
\begin{align}
	\label{eq:evenoddResummedLinesSG}
	\overline G_\mathrm{nl}(x,x_{\nu}) 
	&\equiv\overline G_\mathrm{e}(x,x_{\nu}) - \overline G_\mathrm{o}(x,x_{\nu})
	\nonumber\\
	&= \cosh\left[G(x,x_{\nu})\right] - \sinh\left[G(x,x_{\nu})\right]
	\,,
\end{align}
cf.~\Fig{LinkResummation}(b) for an illustration.

\begin{figure}[t]
	\centering
	\includegraphics[width=0.8\columnwidth]{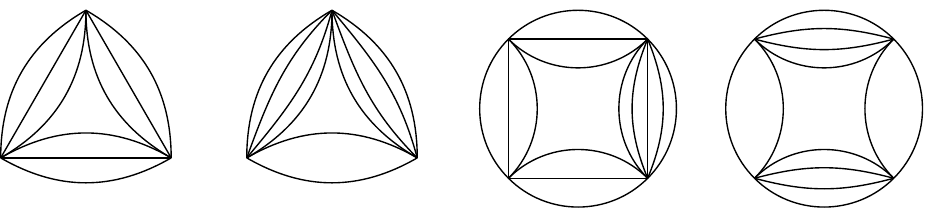}
	\caption{Exemplary diagrams contributing to the $s$-channel bubble-re-summed two-particle irreducible effective action $\Gamma_{2}[G]$, Eqs.~\eq{Gamma2engeq3} and \eq{Gamma2ongeq3}, of the sine-Gordon model.
	Lines represent fully dressed Green's functions $\overline{G}(x,y)$, dots represent $m$-point vertices ${v}_{m}\varphi^{m}$, \Eq{Vexpansion}.
	Diagrams with an even (odd) number of lines linking neighb\bae{our}{or}ing vertices contribute to $\Gamma_{2,\mathrm{e}}[G]$ ($\Gamma_{2,\mathrm{e}}[G]$), cf.~Eqs.~\eq{Gamma2engeq3} and \eq{Gamma2ongeq3}, respectively.}
	\label{fig:RingChainDiagramsSG}
\end{figure}  

\subsection{$s$-channel re-summed 2PI effective action}
\label{sec:Resum2PIEASG}
In the following, we use the re-summed lines $\overline G_\mathrm{l}(x,x)$, $\overline G_\mathrm{e}(x,x_{\nu})$, and $\overline G_\mathrm{o}(x,x_{\nu})$ obtained in the previous subsection to construct the 2PI effective action summing up all ring and chain diagrams including local tadpole loops.
As noted above, the sine-Gordon model, for vanishing mean field $\phi=0$, is invariant under $\varphi\to-\varphi$ and thus gives rise to even-power $\varphi^{2n}$ vertices with $n\in\mathbb{N}$.
Hence, in the ring diagrams, each vertex is connected by either an even or an odd number of propagators $G$ to both its two neighb\bae{our}{or} vertices.
Moreover, if there is an even (odd) number of propagator lines connecting one pair of vertices, all vertex pairs in the ring are connected by an even (odd) number of lines. 
The different possible diagrams are depicted in \Fig{RingChainDiagramsSG}.
Note that in the rings with an odd number of lines linking each vertex pair, there can be only one link with a single line only. 
In that case all other links have a minimum number of three propagators in order for the overall ring diagram to be two-particle irreducible. 
Rings with an even number of lines between the vertices are not subject to such a limit.

To be more consistent one could include also open chains with an arbitrary number of vertices connected to either one or two neighb\bae{our}{or}ing vertices.
These can contribute in contrast to the case of $\phi^{4}$ theory because, if all vertices are linked by three or more lines, also chains are two-particle-irreducible.
We will not include open chains, however, as one can eventually show that, because there is no momentum flowing through them, their effect can be accommodated into a renormal\bae{isation}{ization} of the coupling $\lambda$.
As we are interested in scaling only, neglecting this will not affect our results.

In concrete terms, we need to work out the expectation value \eq{gamma2} of the exponentiated $S_\mathrm{int}$, keeping only 2PI diagrams in terms of bare vertices and (re-summed) full propagator lines.
To begin with, a pre-factor $\mathrm{i}^n/n!$ arises, for the diagrams of order $\lambda^{n}$, from the expansion of $\exp(\mathrm{i} S_{\mathrm{int}})$. 
There are $(n-1)!/2$ ways to arrange $n\geq3$ vertices within a ring, corresponding to the number of their inequivalent circular permutations. 
Hence, each ring diagram containing $n\geq3$ vertices comes with a pre-factor $\mathrm{i}^n/2n$. 

Next we use the re-summed lines of the previous subsection to construct the diagrams shown in \Fig{RingChainDiagramsSG}.
Let us focus, first, on the case of an even number of connecting lines. 
We enumerate the vertices by $j \in \lbrace 1, \ldots, n \rbrace$, implying cyclicity by identifying indices $n+1=1$ and $n=0$. The number of propagators between vertices $j$ and $j+1$ is denoted by $2k_j$.
In addition, $l_{j}$ local tadpole loops are connected to each vertex,  such that the $j$-th vertex is of order $\varphi^{2 (k_{j-1} + k_{j}+l_{j})}$. 
(Note the different meaning of the multiplicities $l_{j}$ and $k_{j}$ as to before.)

Without the need of going into further details, we can now use the arguments of the previous subsection to write down the 2PI effective action containing all rings of $n\geq3$ vertices, connected by an even number of propagator lines, plus the ``daisies'' with an arbitrary high number of petals, which are of order $\lambda$, 
\begin{widetext}
\begin{align}
\label{eq:Gamma2engeq3}	
	\Gamma_{2,\mathrm{e}}[G] 
	&=
	- \sum_{n=1}^{\infty} \frac{\mathrm{i}}{2n} 
	\prod_{j=1}^n \int_{x_j,\mathcal{C}} \mathrm{i}\lambda\, 
	\left.\overline G(x_{j},x_{j})\left[\overline G_\mathrm{e}(x_{j},x_{j+1})-1\right]\,\right|_{x_{n+1}=x_{1}}
	=
	\frac{\mathrm{i}}{2}\Tr\ln_M\left\lbrace \delta_{\mathcal{C}}(x - y) 
	- \mathrm{i}\bar\lambda(x)\left[ \cosh G(x,y) - 1\right]\right\rbrace
	\,,
\end{align}
\end{widetext}
where we dropped the index l of the local re-summed lines $\overline G_\mathrm{l}(x,x)$ and defined the multi-tadpole-dressed local coupling 
\begin{align}
	\label{eq:lambdabar}
	\bar\lambda(x) \equiv \lambda\,\mathrm{e}^{-G(x,x)/2}\,.
\end{align}
The subscript $M$ in $\ln_M$ emphas\bae{ise}{ize}s that the logarithm is understood to be that of a matrix.
The hyperbolic cosine, as before, is a function of the scalar matrix elements $G(x,y)$ of the propagator. 

Note that the sum over $n$ has been deliberately chosen to start at $1$ for the logarithmic-series re-summation to be possible.
The sum thus includes first- and second-order terms in the coupling $\bar\lambda$, which we had not been included in the discussion of the pre-factors for the rings and chains above. 
It is important to real\bae{ise}{ize} that not all possible first- and second-order 2PI diagrams are included in these additional terms, as also at any higher order in $\bar\lambda$, the rings and chains represent only a subclass of all 2PI diagrams.
This is similar to, e.g., an expansion of the 2PI effective action of an O($N$) $\lambda(\varphi_{a}\varphi_{a})^{2}$ model ($a=1,...,N$) in powers of $1/N$, in the limit of a large number $N$ of distinguishable components, where, at any order, the diagrams included represent only a subclass of all 2PI vacuum diagrams at that order in the coupling.
Here, we could also have chosen a different scheme for the re-summation, such as a large-$N$ re-summation similar to that for the quartic theory.
Since it is impossible, though, to formally re-sum all NLO terms in such an approximation, we restrict ourselves to the above quoted 
rings. 
This will be sufficient, though, for the scaling analysis to be done in the following.

Analogously, the contribution including an odd number of lines in each link of the rings, completing again the sums to run from $n=1$, reads
\begin{widetext}
\begin{align}
\label{eq:Gamma2ongeq3}	
	\Gamma_{2,\mathrm{o}}[G]
	&=
	-\ \sum_{n=1}^{\infty} \frac{\mathrm{i}}{2n} 
	\prod_{j=1}^n \int_{x_j,\mathcal{C}} (-\mathrm{i}\lambda)\, 
	\left.\overline G(x_{j},x_{j})\left[\overline G_\mathrm{o}-G\right](x_{j},x_{j+1})\,\right|_{x_{n+1}=x_{1}}
	\nonumber\\
	&\ \quad-
	\frac{\lambda}{2}\sum_{n=1}^{\infty}
	\prod_{j=1}^{n-1} \left\{\int_{x_j,\mathcal{C}}(-\mathrm{i}\lambda) \,
	\overline G(x_{j},x_{j})\left[\overline G_\mathrm{o}-G\right](x_{j},x_{j+1}) \right\}
	\int_{x_n,\mathcal{C}}
	\overline G(x_{n},x_{n})\,
	G(x_n,x_{1})
	\nonumber\\
	&=
	\frac{\mathrm{i}}{2}\Tr\ln_{M}\left\lbrace\delta_{\mathcal{C}}(x-y) 
	+ \mathrm{i} \bar\lambda(x)\left[\sinh G - G\right](x,y)\right\rbrace
	-\frac{1}{2}\int_{x,y,\mathcal{C}}G(x,y)\left[\frac{1}{1+\mathrm{i}\bar\lambda (\sinh G - G)}\right](x,y)\bar\lambda(y)
	\,.
\end{align}
\end{widetext}
Here, the second line, and thus the last term in the third line, includes closed loop chain diagrams with a single propagator link between a single pair of vertices.
A corresponding term, including an open loop chains, could be added to the even part, \Eq{Gamma2engeq3}, but was left out as it leads to subdominant effects only later.

\subsection{Self-energies of the re-summed 2PI effective action}
\label{sec:SelfEnResum2PIEASG}
In a first step towards evaluating the dynamic equations \eq{KB}, one determines the contributions to the self-energy, using the definitions \eq{self_energy_loc_nonloc} and \eq{self_energy_def}.
While the local contribution results as
\begin{align}
\label{eq:SelfEnergyLocal}
	{\Sigma}^{(0)}(x)
	&=
	\lambda \mathrm{e}^{-3G(x,x)/2}&
	\nonumber\\
	&\quad+\ 
	\frac{1}{2}
	\int_{z,\mathcal{C}} 
	\Big[
	\Pi_{\mathrm{e}}(x,z)\Lambda_{\mathrm{e}}(z,x)-\Pi_{\mathrm{o}}(x,z)\Lambda_{\mathrm{o}}(z,x)
	\Big]
	\nonumber\\
	&\quad-\ 
	\frac{1}{2}
	\int_{z,\mathcal{C}}
	G(x,z)\Lambda_{\mathrm{o}}(z,x)
	\,,
\end{align} 
the even and odd non-local self-energies read
\begin{subequations}
\label{eq:SelfEnergies}
\begin{align}
	\overline{\Sigma}_{\mathrm{e}}(x,y)
	&=
	\left[G(x,y)+\Pi_{\mathrm{o}}(x,y)\right]\, 
	I_{\mathrm{e}}(y,x)
	\,,\\
	\overline{\Sigma}_{\mathrm{o}}(x,y)
	&=
	\left[1+\Pi_{\mathrm{e}}(x,y)\right]\,I_{\mathrm{o}}(y,x)
	\,,
\end{align} 
\end{subequations}
where the multiloop functions summing up the multiple-propagator links are defined as
\begin{subequations}
\label{eq:LoopFcts}
\begin{align}
	\Pi_{\mathrm{e}}(x,y)
	&=\cosh G(x,y) - 1
	\,,\\
	\Pi_{\mathrm{o}}(x,y)
	&=
	\sinh G(x,y)-G(x,y)
	\,.
\end{align} 
\end{subequations}
The chains constructed out of lining up these loops are encoded in
\begin{subequations}
\label{eq:IntegralsIJH}
\begin{align}
	I_{\mathrm{e}}(x,y)
	&=
	-\int_{z,\mathcal{C}} 
	\bar\lambda(x)\,\Pi_{\mathrm{e}}(x,z)\,
	\Lambda_\mathrm{e}(z,y)
	\,,
	\\ 
	I_{\mathrm{o}}(x,y)
	&=
	-\int_{z,\mathcal{C}} 
	\bar\lambda(x)\,\Pi_{\mathrm{o}}(x,z)\,
	\Lambda_\mathrm{o}(z,y)
	\,,
\end{align}
\end{subequations}
with non-perturbative non-local coupling functions
\begin{subequations}
\begin{align}
	\label{eq:Lambdae}
	\Lambda_{\mathrm{e}}(x,y)
	&= \left[{1-\mathrm{i}\bar\lambda\, \Pi_{\mathrm{e}}}\right]^{-1}\!\!(x,y)\,\bar\lambda(y)
	\,,
	\\
	\label{eq:Lambdao}
	\Lambda_{\mathrm{o}}(x,y)
	&= \left[{1+\mathrm{i}\bar\lambda\, \Pi_{\mathrm{o}}}\right]^{-1}\!\!(x,y)\,\bar\lambda(y)
	\,,
\end{align}
\end{subequations}
where the inverse is to be understood, as in the 2PI effective action, as the matrix inverse.
Note that all of the above functions are symmetric under exchange of their arguments, i.e., $G(x,y)=G(y,x)$ implies $\Pi_{\mathrm{e/o}}(x,y)=\Pi_{\mathrm{e/o}}(y,x)$, $I_{\mathrm{e/o}}(x,y)=I_{\mathrm{e/o}}(y,x)$, etc.
We furthermore remark that we have left out contributions to the odd self energies \eq{SelfEnergies}, which arise from taking the derivative of the term in square brackets in the second contribution to \eq{Gamma2ongeq3}, as they will be subdominant later.

\subsection{Decomposition of the self-energies}
\label{sec:KBEq}
To arrive at the Kadanoff-Baym equations \eq{KB}, we need to decompose the non-local self-energies according to \Eq{self_energy_decomp}. 
Applying, to $\cosh G(x,y)$, $\sinh G(x,y)$, the decomposition \eq{G_decomp} of $G$ and standard trigonometric identities 
we can decompose the loop functions \eq{LoopFcts} into their spectral and statistical components
\begin{subequations}
	\label{eq:Pi}
	\begin{align}
		\label{eq:Pi_F_e}
		\Pi^F_{\mathrm{e}}
		&=
		\cosh F \cos(\rho/2) - 1 
		\,,&
		\Pi^{\,\rho}_{\mathrm{e}}
		&=
		2\sinh F \sin(\rho/2)\,,
		\\
		\label{eq:Pi_F_o}	
		\Pi^F_{\mathrm{o}}
		&=
		\sinh F \cos(\rho/2) - F
		\,,&
		\Pi^{\,\rho}_{\mathrm{o}}
		&=
		2\cosh F\sin(\rho/2)-\rho\,,
	\end{align}
\end{subequations}
where we suppressed the identical arguments $(x,y)$ of all (products of) functions.
These functions, in particular the retarded and advanced loops 
\begin{subequations}
	\label{eq:PieoRA}
\begin{align}
	\label{eq:PieoR}
  	\Pi_{\mathrm{e/o}}^R(x,y) 
  	& =  \theta(x_{0}-y_{0})\, \Pi_{\mathrm{e/o}}^{\,\rho}(x,y)
	\,,
  	\\
	\label{eq:PieoA}
  	\Pi_{\mathrm{e/o}}^A(x,y) 
  	& =  -\theta(y_{0}-x_{0})\, \Pi_{\mathrm{e/o}}^{\,\rho}(x,y)
	 \,,
\end{align}
\end{subequations}
are an essential ingredient to the non-perturbative integrals $I$,
\eq{IntegralsIJH}, which can also be decomposed into statistical and spectral components. 
For this, it is useful to first re-write the integrals in terms of implicit integral equations and after decomposing these, re-express them in terms of non-perturbative coupling functions. 

For the explicit steps, see \App{KBEq}. 
As a result of these, the statistical and spectral components of the integrals \eq{IntegralsIJH} can be written as
\begin{subequations}
\label{eq:IntegralsIJHFrho}
\begin{align}
	\label{eq:IeoF}
	I^{F}_{\mathrm{e/o}}
	&=
	-\Lambda^{R}_{\mathrm{e/o}}\ast
	\Pi^{F}_{\mathrm{e/o}}
	\ast\Lambda^{A}_{\mathrm{e/o}}
	\,,
	\\ 
	\label{eq:Ieorho}
	I^{\,\rho}_{\mathrm{e/o}}
	&=
	-\Lambda^{R}_{\mathrm{e/o}}\ast
	\Pi^{\,\rho}_{\mathrm{e/o}}
	\ast\Lambda^{A}_{\mathrm{e/o}}
	\,,
\end{align} 
\end{subequations}
with the non-perturbative couplings
\begin{subequations}
\label{eq:LambdaeoRA}
\begin{align}
	\Lambda^{R}_{\mathrm{e/o}}
	&= \left(1\mp \bar\lambda\ast\Pi^{R}_\mathrm{e/o}\right)^{-1}\!\!\ast\bar\lambda
	\,,
	\\
	\Lambda^{A}_{\mathrm{e/o}}
	&= \left(1\mp \bar\lambda\ast\Pi^{A}_\mathrm{e/o}\right)^{-1}\!\!\ast\bar\lambda
	\,.
\end{align}
\end{subequations}
where the star $\ast$ serves as a short-hand notation for matrix products in space-time, cf.~\App{notation}, $\bar\lambda\equiv\bar\lambda(x)\,\delta(x-y)$ is a diagonal matrix.

The above integrals are finally needed in the statistical and spectral components of the self-energies \eq{SelfEnergies},
\begin{subequations}
	\label{eq:SelfEnergiesDecomposed}
	\begin{align}
		\label{eq:self_energy_e_F}
		{\Sigma}^{F}_{\mathrm{e}}
		&=
		\left(F+\Pi^{F}_{\mathrm{o}}\right)\cdot I^{F}_{\mathrm{e}}
		-\left(\rho+\Pi^{\,\rho}_{\mathrm{o}}\right)\cdot I^{\,\rho}_{\mathrm{e}}/4
		\,,\\ 
		\label{eq:self_energy_e_rho}
		{\Sigma}^{\,\rho}_{\mathrm{e}}
		&=
		\left(\rho+\Pi^{\,\rho}_{\mathrm{o}}\right)\cdot  I^{F}_{\mathrm{e}}
		+\left(F+\Pi^{F}_{\mathrm{o}}\right)\cdot I^{\,\rho}_{\mathrm{e}}
		\,,\\ 
		\label{eq:self_energy_o_F}
		{\Sigma}^{F}_{\mathrm{o}}
		&=
		\left(1+\Pi^{F}_{\mathrm{e}}\right)\cdot I^{F}_{\mathrm{o}}
		-\Pi^{\,\rho}_{\mathrm{e}}\cdot 
		I^{\,\rho}_{\mathrm{o}}
		/4
		\,,\\ 
		\label{eq:self_energy_o_rho}
		{\Sigma}^{\,\rho}_{\mathrm{o}}
		&=
		-\left(1+\Pi^{F}_{\mathrm{e}}\right)\cdot 
		I^{\,\rho}_{\mathrm{o}}
		-\Pi^{\,\rho}_{\mathrm{e}}\cdot 
		I^{F}_{\mathrm{o}}
		\,,
\end{align} 
\end{subequations}
where we use the short-hand notation summar\bae{ise}{ize}d in \App{notation} for products of functions in position space, taking into account that spectral components are odd under an exchange of their coordinates $x$ and $y$.
Note that $\Sigma^{(0)}$ is local in space-time and thus is equal to its statistical part.
The above self-energy components are now ready for use in the Kadanoff-Baym dynamic equations \eq{KB}, where the even and odd contributions are to be summed up, $\Sigma^{F}=\Sigma^{F}_{\mathrm{e}}+\Sigma^{F}_{\mathrm{o}}$, $\Sigma^{\,\rho}=\Sigma^{\,\rho}_{\mathrm{e}}+\Sigma^{\,\rho}_{\mathrm{o}}$.

The mass shift, cf.~\eq{RenMass}, finally reads
\begin{align}
		\label{eq:self_energy_0}
		{\Sigma}^{(0)}
		=&\Bigg\{\lambda\exp\left[-3F/2\right]
		+\frac{1}{2}\left[
		I^{F}_{\mathrm{o}}-I^{F}_{\mathrm{e}}
		\right]\lambda^{-1}\exp\left[F/2\right]
		\nonumber\\
		&\quad-\
		\frac{1}{2}
		\Lambda^{R}_{\mathrm{o}}\ast F\ast \Lambda^{A}_{\mathrm{o}}
		\Bigg\}
		(x,x)
		\,.
\end{align} 
The resulting dynamic equations describe the evolution according to in general non-perturbatively strong correlations built up through the sine-Gordon interactions.
The form of the coupling functions \eq{LambdaeoRA} demonstrates that, as is typical for loop-chain re-summations, for the dynamics to be non-perturbative, the loop function multiplied with the (dressed) local coupling must, at least, be of order unity, $|\bar\lambda\ast\Pi^{\,\rho}_{\mathrm{e/o}}|\simeq1$.
We will later be interested in the limit $|\bar\lambda\ast\Pi^{\,\rho}_{\mathrm{e/o}}|\gg1$, in which the dressed local coupling $\bar\lambda$ drops out, and the non-perturbative couplings scale as $\Lambda^{R,A}_{\mathrm{e/o}}\sim\big(\Pi^{R,A}_{\mathrm{e/o}}\big)^{-1}$ alone.
This will be important for the scaling analysis close to a non-thermal fixed point where a strong infrared mode occupancy  allows this limit to be reached.

Taking into account non-linear interactions, the quasiparticle spectral function is modified in general, and quasiparticles can acquire a finite lifetime. 
The dispersion relation may then significantly differ from its free form $\omega^{(0)}_{\mathbf{p}}$. 
Within the quasiparticle picture introduced above  and the considered spatially uniform system, the dressed coupling \eq{lambdabar} evaluates to
\begin{align}
	\label{eq:lambdabarf}
	\bar\lambda(x) 
	&= \lambda\exp\left[-F(x,x)/2\right] = \lambda\exp\left[-F(X,s=0)/2\right] 
	\nonumber\\
	&
	=\lambda\exp\left\{-\frac{\eta}{2}\int_{\mathbf{p}}\frac{f(t,\mathbf{p})+1/2}{\omega_{\mathbf{p}}}\right\}
	\,,
\end{align}
where we used the symmetry \eq{FrhoSymmetries}, and assumed the free spectral function \eq{rho_free}, as well as a parity-even distribution, $f(t,\mathbf{p})=f(t,-\mathbf{p})$.
As is usually encountered for such loop integrals, the exponent is divergent unless one specifies a UV cutoff or introduces a  renormal\bae{isation}{ization} of the bare couplings $\eta$ and $\lambda$. 
Moreover, the integral over the quasiparticle distribution must converge and thus requires a physical cutoff or sufficiently fast decay of $f$ for large momenta.
In the non-relativistic approximation considered here, we assume a cutoff on the order of $M$ to regular\bae{ise}{ize} all momentum integrals and write the coupling constant as $\bar\lambda\equiv\lambda\exp\left(-\eta n_{0}/2M\right)$, in terms of the total quasiparticle density 
\begin{align}
	\label{eq:n0QP}
	n_{0} 
	= \int_\mathbf{p}f(t,\mathbf{p})\,.
\end{align}
During the scaling evolutions considered in the following, this density is anticipated to be conserved in time, $n_{0}(t)\equiv n_{0}$.
In that case, we can pull the couplings $\bar\lambda$, in the integrals \eq{IntegralsIJHFrho} and couplings \eq{LambdaeoRA}, as constant factors to the front, removing thereby all multiplications with the unit matrix $\delta(x-y)$.

\subsection{Scattering Integral}
\label{sec:ScattInt}
To arrive at an explicit expression of the kinetic equation, it remains to Fourier transform the self-energies \eq{SelfEnergiesDecomposed} entering the scattering integral \eq{Cf} with respect to the relative coordinates $s$ and insert the quasiparticle parametr\bae{isation}{ization} of the statistical and spectral functions as introduced in the previous section. 
As pointed out, we consider an on average spatially uniform system, taking $\bar\lambda$ as a renormal\bae{ise}{ize}d coupling constant.
As a consequence of this, the integrals \eq{IntegralsIJHFrho}, in Fourier space, to the lowest-order of a gradient expansion, can be written as 
\begin{subequations}
\label{eq:IntegralsIJHFrhoFourier}
\begin{align}
	\label{eq:IeoFp}
	I^{F}_{\mathrm{e/o}}(t,p)
	&=
	- \left(\left|\Lambda^{R}_{\mathrm{e/o}}\right|^{2}\cdot\Pi^{F}_{\mathrm{e/o}}\right)(t,p)
	\,,\\ 
	\label{eq:Ieorhop}
	I^{\,\rho}_{\mathrm{e/o}}(t,p)
	&=
	- \left(\left|\Lambda^{R}_{\mathrm{e/o}}\right|^{2}\cdot\Pi^{\,\rho}_{\mathrm{e/o}}\right)(t,p)
	\,,
\end{align} 
\end{subequations}
where the dot denotes a product in momentum space, cf.~\App{notation}. 

The effective momentum-dependent coupling function is defined as
\begin{align}
	\label{eq:LambdaReopMain}
	\Lambda^{R}_{\mathrm{e/o}}(t,p)
	&=\frac{\bar\lambda}{1 \mp \bar\lambda\,\Pi_{\mathrm{e/o}}^R(t,p)}
	\,,
\end{align}	
where we used the symmetry $\Pi^{A}_{\mathrm{e/o}}(t,p)=\Pi^{R}_{\mathrm{e/o}}(t,-p)=\big[\Pi^{R}_{\mathrm{e/o}}(t,p)\big]^{\ast}$.
The above momentum-dependent integrals convolved with correlation and loop functions determine the self-energies in Fourier space, obtained from Eqs.~\eq{SelfEnergiesDecomposed} by replacing dots with stars,
\begin{subequations}
	\label{eq:NonlocSelfEnergiesofp}
	\begin{align}
		\label{eq:self_energy_e_F_p}
		{\Sigma}^{F}_{\mathrm{e}}
		&=
		\left(F+\Pi^{F}_{\mathrm{o}}\right)\ast I^{F}_{\mathrm{e}}
		-\left(\rho+\Pi^{\,\rho}_{\mathrm{o}}\right)\ast I^{\,\rho}_{\mathrm{e}}/4
		\,,\\ 
		\label{eq:self_energy_e_rho_p}
		{\Sigma}^{\,\rho}_{\mathrm{e}}
		&=
		\left(\rho+\Pi^{\,\rho}_{\mathrm{o}}\right)\ast  I^{F}_{\mathrm{e}}
		+\left(F+\Pi^{F}_{\mathrm{o}}\right)\ast I^{\,\rho}_{\mathrm{e}}
		\,,\\ 
		\label{eq:self_energy_o_F_p}
		{\Sigma}^{F}_{\mathrm{o}}
		&=
		\left(1+\Pi^{F}_{\mathrm{e}}\right)\ast I^{F}_{\mathrm{o}}
		-\Pi^{\,\rho}_{\mathrm{e}}\ast 
		I^{\,\rho}_{\mathrm{o}}/4
		\,,\\ 
		\label{eq:self_energy_o_rho_p}
		{\Sigma}^{\,\rho}_{\mathrm{o}}
		&=
		-\left(1+\Pi^{F}_{\mathrm{e}}\right)\ast 
		I^{\,\rho}_{\mathrm{o}}
		-\Pi^{\,\rho}_{\mathrm{e}}\ast 
		I^{F}_{\mathrm{o}}
		\,.
	\end{align} 
\end{subequations}
For expressing the scattering integral \eq{Cf} in terms of the quasiparticle mode occupations $f(t,\mathbf{p})$ it is convenient to rewrite it in terms of the `greater' and `smaller' components of the correlators and self-energies, which, by use of \Eq{FDR}, are 
\begin{subequations}
\label{eq:Ggtrless}
\begin{align}
	G^{>}(t,p) 
	&= F(t,p)+\mbox{$\frac{\mathrm{i}}{2}$}\rho(t,p)
	= -\mathrm{i}f(t,p)\,\rho(t,p)
	\,,\\
	G^{<}(t,p) 
	&= F(t,p)-\mbox{$\frac{\mathrm{i}}{2}$}\rho(t,p)
	= -\mathrm{i}\left[f(t,p)+1\right]\,\rho(t,p)
	\,
\end{align}
\end{subequations}
and depend on central time $t=(x_{0}+y_{0})/2$ and 4-momentum $p$.
The statistical and spectral components of the self-energies and integrals $I$ are combined analogously.
In terms of these combinations, the self-energies \eq{NonlocSelfEnergiesofp} read
\begin{align}
	\label{eq:NonlocSelfEnergiesGrtSm}
	{\Sigma}^{\gtrless}_{\mathrm{e}}
	&=
	\left(G^{\gtrless}+\Pi^{\gtrless}_{\mathrm{o}}\right)\ast I^{\gtrless}_{\mathrm{e}}
	\,,& 
	{\Sigma}^{\gtrless}_{\mathrm{o}}
	&=
	\left(1+\Pi^{\lessgtr}_{\mathrm{e}}\right)\ast I^{\lessgtr}_{\mathrm{o}}
	\,,
\end{align} 
with the loop functions obtained from Eqs.~\eq{Pi},
\begin{align}
	\label{eq:Pigrsm}
	\Pi^{\gtrless}_{\mathrm{e}}
	&=
	\cosh G^{\gtrless} - 1 
	\,,\quad
	\Pi^{\gtrless}_{\mathrm{o}}
	=
	\sinh G^{\gtrless}  - G^{\gtrless}
	\,,
\end{align}
and $I$-integrals
\begin{align}
	\label{eq:Ieo_gtrless_p}
	I^{\gtrless}_{\mathrm{e/o}}(t,p)
	&=
	- \left(\left|\Lambda^{R}_{\mathrm{e/o}}\right|^{2}\cdot\Pi^{\gtrless}_{\mathrm{e/o}}\right)(t,p)
	\,.
\end{align}
With these, the integrand of \eq{Cf} results as
\begin{align}
	\label{eq:CfIntegrand}
	&-\mathrm{i}\left[
	\tildeft{\Sigma}^{\,\rho}_{\mathrm{e/o}}(t,p) F(t,p) - \Sigma^F_{\mathrm{e/o}}(t,p)\,\tildeft{\rho}(t,p)
	\right]
	\nonumber\\
	&\qquad
	=
	\left(\Sigma^{<}_{\mathrm{e/o}}\cdot G^{>} - \Sigma^{>}_{\mathrm{e/o}}\cdot G^{<}\right)(t,p)
	\,,
\end{align} 
and, inserting Eqs.~\eq{NonlocSelfEnergiesGrtSm}--\eq{Pigrsm}, one arrives at the kernels
\begin{align}
	\label{eq:CfIntegrand_e}
	&-\mathrm{i}\left[
	\tildeft{\Sigma}^{\,\rho}_{\mathrm{e}}\cdot F - \Sigma^F_{\mathrm{e}}\cdot\tildeft{\rho}
	\right](t,p)
	\nonumber\\
	&\quad
	=
	\left\{\left(\sinh G^{>}\ast\left[\left|\Lambda^{R}_{\mathrm{e}}\right|^{2}\cdot\left(\cosh G^{>} - 1\right)\right]\right)
	\cdot G^{<} \right.
	\nonumber\\
	&\quad\ \ -\
	\left.\left(\sinh G^{<}\ast\left[\left|\Lambda^{R}_{\mathrm{e}}\right|^{2}\cdot\left(\cosh G^{<} - 1\right)\right]\right)
	\cdot G^{>}\right\}(t,p)
	\,,\\
	\label{eq:CfIntegrand_o}
	&-\mathrm{i}\left[
	\tildeft{\Sigma}^{\,\rho}_{\mathrm{o}}\cdot F - \Sigma^F_{\mathrm{o}}\cdot\tildeft{\rho}
	\right](t,p)
	\nonumber\\
	&\quad
	=\left\{\left(\cosh G^{>}\ast\left[\left|\Lambda^{R}_{\mathrm{o}}\right|^{2}\cdot\left(\sinh G^{>} - G^{>}\right)\right]\right)
	\cdot G^{<} \right.
	\nonumber\\
	&\quad\ \ -\
	\left.\left(\cosh G^{<}\ast\left[\left|\Lambda^{R}_{\mathrm{o}}\right|^{2}\cdot\left(\sinh G^{<} - G^{<}\right)\right]\right)
	\cdot G^{>}\right\}(t,p)
	\,.
\end{align} 
\begin{widetext}
Inserting these into \Eq{Cf} gives the scattering integral
\begin{align}
	\label{eq:Cfmain}
	C[f](t,\mathbf{p})
	=
	\int_0^{\infty} &\frac{\dd{p^0}}{2\pi} \,
	\Bigg(
	\left\{\left[\left|\Lambda^{R}_{\mathrm{e}}\right|^{2}\cdot\left(\cosh G^{>} - 1\right)\right]\ast\sinh G^{>}\right.
	+\left.\left[\left|\Lambda^{R}_{\mathrm{o}}\right|^{2}\cdot\left(\sinh G^{>} - G^{>}\right)\right]\ast\cosh G^{>}\right\}
	\cdot G^{<} 
	\nonumber\\
	&\quad\,-\
	\left\{\left[\left|\Lambda^{R}_{\mathrm{e}}\right|^{2}\cdot\left(\cosh G^{<} - 1\right)\right]\ast\sinh G^{<}\right.
	+\left.\left[\left|\Lambda^{R}_{\mathrm{o}}\right|^{2}\cdot\left(\sinh G^{<} - G^{<}\right)\right]\ast\cosh G^{<}\right\}
	\cdot G^{>}\Bigg)(t,p)
	\,.
\end{align}
Note that the hyperbolic functions are defined in the convolutional sense, $\cosh G = 1 + G\ast G/2 + \dots$, etc. 
In the next step, we expand out the hyperbolic functions in \eq{Cfmain}, using \eq{Ggtrless}, to obtain the scattering integral
\begin{align}
	\label{eq:Cfsumneo}
	&C[f](t,\mathbf{p})
	=
	\sum_{n=1}^{\infty}\Big[C^{(n)}_{\mathrm{e}}[f](t,\mathbf{p})+C^{(n)}_{\mathrm{o}}[f](t,\mathbf{p})\Big]
	\,,
\end{align}
with the even and odd contributions of $n$th order in the expansion in powers of $f$,
\begin{subequations}
\begin{align}
	\label{eq:Cfen-wFreqInt}
	C^{(n)}_{\mathrm{e}}[f](t,\mathbf{p})
	=\sum_{m=1}^{n}
	\int_0^{\infty} \frac{\dd{p^0}}{2\pi} 
	\int_{q_{1}\dots q_{2n+1}}\!
	&\frac{\delta(p-q_{1}-\dots-q_{2n+1})}{(2m)!(2[n-m]+1)!}
	\left|\Lambda^{R}_{\mathrm{e}}(q_{1}+\dots+q_{2m})\right|^{2}\,(-1)^{n}\rho(p)\,\rho(q_{1})\cdots\rho(q_{2n+1})
	\nonumber\\
	&\quad\times\
	\left[(f_{q_{1}}+1)\cdots(f_{q_{2n+1}}+1)f_{p}-f_{q_{1}}\cdots f_{q_{2n+1}}(f_{p}+1)\right]
	\,,\\
	\label{eq:Cfon-wFreqInt}
	C^{(n)}_{\mathrm{o}}[f](t,\mathbf{p})
	=\sum_{m=1}^{n}
	\int_0^{\infty} \frac{\dd{p^0}}{2\pi} 
	\int_{q_{1}\dots q_{2n+1}}\!
	&\frac{\delta(p-q_{1}-\dots-q_{2n+1})}{(2m+1)!(2[n-m])!}
	\left|\Lambda^{R}_{\mathrm{o}}(q_{1}+\dots+q_{2m+1})\right|^{2}\,(-1)^{n}\rho(p)\,\rho(q_{1})\cdots\rho(q_{2n+1})
	\nonumber\\
	&\quad\times\
	\left[(f_{q_{1}}+1)\cdots(f_{q_{2n+1}}+1)f_{p}-f_{q_{1}}\cdots f_{q_{2n+1}}(f_{p}+1)\right]
	\,.
\end{align}
\end{subequations}
\end{widetext}
Here we have suppressed the dependence on the time $t$. 
Integrating out the frequencies $p^{0}$, $q^{0}_{i}$ is straightforwardly possible for the free spectral function \eq{rho_free}, which leads to the final form of $C[f]$, as a sum of multidimensional integrals over the three momenta, involving a scattering `$T$-matrix' ,energy- and momentum-conservation constraints, and a sum of in- and out-scattering terms depending on  the quasiparticle distribution $f(t,\mathbf{p})$ only, 
\begin{align}
	\label{eq:Cfmain2}
	&C[f](t,\mathbf{p})
	\equiv
	\sum_{n=1}^{\infty}C^{(n)}[f](t,\mathbf{p})
	=-\sum_{n=1}^{\infty}
	\int\prod_{i=1}^{2n+1}\frac{\dd{\mathbf{q}_{i}}}{(2\pi)^{d}} \,
	\nonumber\\
	&\quad\times\
	\left|T^{(n)}(t;{\mathbf{p},\mathbf{q}_{1},\dots,\mathbf{q}_{2n+1}})\right|^{2}
	\nonumber\\
	&\quad\times\
	\delta(\omega_{\mathbf{p}}-\omega_{\mathbf{q}_{1}}-\dots-\omega_{\mathbf{q}_{n+1}}
	+\omega_{\mathbf{q}_{n+2}}+\dots+\omega_{\mathbf{q}_{2n+1}})\,
	\nonumber\\
	&\quad\times\
	\delta(\mathbf{p}-\mathbf{q}_{1}-\dots-\mathbf{q}_{n+1}+\mathbf{q}_{n+2}+\dots+\mathbf{q}_{2n+1})
	\nonumber\\
	&\quad\times\
	\left[(f_{\mathbf{q}_{1}}+1)\cdots(f_{\mathbf{q}_{n+1}}+1)f_{\mathbf{q}_{n+2}}\cdots f_{\mathbf{q}_{2n+1}}f_{\mathbf{p}}
	\right.
	\nonumber\\
	&\quad\quad-\
	\left.
	f_{\mathbf{q}_{1}}\cdots f_{\mathbf{q}_{n+1}}(f_{\mathbf{q}_{n+2}}+1)\cdots 
	(f_{\mathbf{q}_{2n+1}}+1)(f_{\mathbf{p}}+1)\right]
	\,.
\end{align}
Here we have suppressed, on the right-hand-side, the dependence of $f_{\mathbf{q}}\equiv f(t,\mathbf{q})$ on the time $t$.
As we are eventually interested in universal transport into the infrared region of small wave numbers, where the dispersion is in general gapped, $\omega(|{\mathbf{p}}|\to0)\to M$ we explicitly write out only the on-energy-shell terms.
Hence, the above scattering integral describes $(n+1)$-to-$(n+1)$ processes for which the sum of all frequencies, $p^{0}+\sum_{i=1}^{2n+1}q^{0}_{i}$ is gapless, i.e., $n+1$ of the frequencies are evaluated in the positive domain, $q^{0}_{i}=\omega(\mathbf{q}_{i})$, $i=1,\dots,n+1$, and a further $n+1$ in the negative domain, $q^{0}_{i}=-\omega(\mathbf{q}_{i})$, $i=n+2,\dots,2n+1$, as well as $p_{0}=-\omega(\mathbf{p})$.

\subsection{Non-perturbative $T$-matrix}
\label{sec:TMatrix}
The $T$-matrices squared, which measure the probability of the multi-momentum scattering processes described by \eq{Cfmain2} are defined as
\begin{align}
	\label{eq:Tmatrix}
	\left|T^{(n)}(t;{\mathbf{p},\mathbf{q}_{1},\dots,\mathbf{q}_{2n+1}})\right|^{2}
	&=\frac{g_{\mathrm{eff}}^{2}(n;t;\mathbf{p},\{\mathbf{q}_{i}\})}{n!(n+1)!}
	\frac{\eta^{2n+2}}{2\omega_{\mathbf{p}}}\prod_{i=1}^{2n+1}\frac{1}{2\omega_{\mathbf{q}_{i}}}
	\,,
\end{align}
in terms of the nonperturbative couplings \eq{LambdaReopMain} entering a (dimensionful) coupling $g_{\mathrm{eff}}$,
\begin{align}
	\label{eq:SM:geff}
	&g_{\mathrm{eff}}^{2}(n;t;\mathbf{p},\{\mathbf{q}_{i}\})
	=\sum_{m=1}^{n}\left[
	\sum_{\{\sigma\}}\left|\Lambda^{R}_{\mathrm{e}}
	\left(t,
	\sum_{i=1}^{2m}s_{\sigma_{i}}\,\omega_{\mathbf{q}_{\sigma_{i}}},
	\sum_{i=1}^{2m}s_{\sigma_{i}}\,\mathbf{q}_{\sigma_{i}}
	\right)\right|^{2}\,\right.
	\nonumber\\
	&\quad\quad+\
	\left.
	\sum_{\{\sigma\}}\left|\Lambda^{R}_{\mathrm{o}}
	\left(t,
	\sum_{i=1}^{2m+1}s_{\sigma_{i}}\,\omega_{\mathbf{q}_{\sigma_{i}}},
	\sum_{i=1}^{2m+1}s_{\sigma_{i}}\,\mathbf{q}_{\sigma_{i}}
	\right)\right|^{2}\,\right]
	\,.
\end{align}
Here, $s_{k}=\mathrm{sgn}(n+3/2-k)$, which is $s_{k}=+1$ if $k\leq n+1$ and $s_{k}=-1$ for $k>n+1$, and the sums over $\sigma\subset\{1,\dots,2n+1\}$ are those over all subsets of $2m$ (or $2m+1$, in the odd case) momenta of all the $\mathbf{q}_{k}$ in a given term.
Note that, to arrive at \Eq{Tmatrix}, we first symmetr\bae{ise}{ize}d over all momenta, i.e., wrote the even-case scattering integral as a sum where every choice of $2m$ momenta out of $2n+1$ momenta is real\bae{ise}{ize}d once and divide by the number $(2n+1)!/[(2m)!(2n+1-2m)!]$ of terms in the sum. 
After that we are free to choose an arbitrary set of $n$ momenta out of $2n+1$ which we evaluate at a negative frequency $-\omega_{\mathbf{q}_{i}}$, if in turn we multiply by the $(2n+1)!/[n!(n+1)!]$ possibilities to do so. 
Combining the above steps yields the combinatorial factor $1/[n!(n+1)!]$.  
The odd case is analogous.

Finally, the loop functions $\Pi^R_{\mathrm{e/o}}$ entering the coupling functions $\Lambda^{R}_{\mathrm{e/o}}$ can be expressed, using \Eq{PieoR}, in terms of $\Pi^{\,\rho}_{\mathrm{e/o}}=-\mathrm{i}(\Pi^{>}_{\mathrm{e/o}}-\Pi^{<}_{\mathrm{e/o}})$.
Using Eqs.~\eq{Pi}, \eq{Pigrsm} they can be expressed as
\begin{subequations} 
    	\label{eq:PiRdefAppG}
\begin{align}
	&\Pi^R_{\mathrm{e}}(t,p^{0},\vec p) 
	= -\mathrm{i}\left[\theta\ast\left(\cosh G^{>}-\cosh G^{<}\right)\right](t,p^{0},\vec p) 
  	\nonumber \\ &
	= \int \frac{\mathrm{d}q^{0}}{2\pi} \frac{1 }{q^{0}+\mathrm{i}\epsilon} 
 	\left(\cosh G^{>}-\cosh G^{<}\right)(t,p^{0}-q^{0},\vec p )\,
    	\,,
    	\\
  	&\Pi^R_{\mathrm{o}}(t,p^{0},\vec p) 
  	= -\mathrm{i}\left[\theta\ast\left(\sinh G^{>}-\sinh G^{<}-G^{>}+ G^{<}\right)\right](t,p) 
	\nonumber \\ &
  	= \int \frac{\mathrm{d}q^{0}}{2\pi} \frac{1 }{q^{0}+\mathrm{i}\epsilon} 
    	\left(\sinh G^{>}-\sinh G^{<}-\mathrm{i}\rho\right)(t,p^{0}-q^{0},\vec p )\,
    	\,,
\end{align}
\end{subequations} 
and, as done for the scattering integral itself, expanded in powers of the distribution function $f$,
\begin{subequations} 
\begin{align}
 	\label{eq:PiReExpanded}
	&\Pi^R_{\mathrm{e}}(t,p^{0},\vec p) 
  	= -\mathrm{i}\left[\theta\ast\left(\cosh G^{>}-\cosh G^{<}\right)\right](t,p^{0},\vec p) 
  	\nonumber \\ &
  	=\sum_{n=1}^{\infty}
  	\int \frac{\mathrm{d}q^{0}}{2\pi} \!\frac{\eta^{2n}}{q^{0}+\mathrm{i}\epsilon}
	\int\prod_{i=1}^{2n}\frac{\dd{\mathbf{q}_{i}}}{(2\pi)^{d}2\omega(\mathbf{q}_{i})} \,
	\sum_{m=0}^{2n}
	\frac{-1}{m!(2n-m)!}
	\nonumber\\
	&\quad\times\
	\delta(p^{0}-q^{0}-\omega_{\mathbf{q}_{1}}-\dots-\omega_{\mathbf{q}_{m}}
	+\omega_{\mathbf{q}_{m+1}}+\dots+\omega_{\mathbf{q}_{2n}})\,
	\nonumber\\
	&\quad\times\
	\delta(\mathbf{p}-\mathbf{q}_{1}-\dots-\mathbf{q}_{m}+\mathbf{q}_{m+1}+\dots+\mathbf{q}_{2n})
	\nonumber\\
	&\quad\times\
	\left[(f_{\mathbf{q}_{1}}+1)\cdots(f_{\mathbf{q}_{m}}+1)f_{\mathbf{q}_{m+1}}\cdots f_{\mathbf{q}_{2n}}
	\right.
	\nonumber\\
	&\quad\quad-\
	\left.
	f_{\mathbf{q}_{1}}\cdots f_{\mathbf{q}_{m}}(f_{\mathbf{q}_{m+1}}+1)\cdots 
	(f_{\mathbf{q}_{2n}}+1)\right]
	\,,\\ 
 	\label{eq:PiRoExpanded}
  	&\Pi^R_{\mathrm{o}}(t,p^{0},\vec p) 
  	= -\mathrm{i}\left[\theta\ast\left(\sinh G^{>}-\sinh G^{<}-G^{>}+ G^{<}\right)\right](t,p) 
  	\nonumber \\ &
  	=\sum_{n=1}^{\infty}
  	\int \frac{\mathrm{d}q^{0}}{2\pi} \!\frac{\eta^{2n+1}}{q^{0}+\mathrm{i}\epsilon}
	\int\prod_{i=1}^{2n+1}\frac{\dd{\mathbf{q}_{i}}}{(2\pi)^{d}2\omega_{\mathbf{q}_{i}}} \,
	\sum_{m=0}^{2n+1}
	\frac{-1}{m!(2n-m+1)!}
	\nonumber\\
	&\quad\times\
	\delta(p^{0}-q^{0}-\omega_{\mathbf{q}_{1}}-\dots-\omega_{\mathbf{q}_{m}}
	+\omega_{\mathbf{q}_{m+1}}+\dots+\omega_{\mathbf{q}_{2n+1}})\,
	\nonumber\\
	&\quad\times\
	\delta(\mathbf{p}-\mathbf{q}_{1}-\dots-\mathbf{q}_{m}+\mathbf{q}_{m+1}+\dots+\mathbf{q}_{2n+1})
	\nonumber\\
	&\quad\times\
	\left[(f_{\mathbf{q}_{1}}+1)\cdots(f_{\mathbf{q}_{m}}+1)f_{\mathbf{q}_{m+1}}\cdots f_{\mathbf{q}_{2n+1}}
	\right.
	\nonumber\\
	&\quad\quad-\
	\left.
	f_{\mathbf{q}_{1}}\cdots f_{\mathbf{q}_{m}}(f_{\mathbf{q}_{m+1}}+1)\cdots 
	(f_{\mathbf{q}_{2n+1}}+1)\right]
  	  \,.
\end{align}
\end{subequations} 
Integrating again over the frequencies gives the final expressions 
\begin{subequations} 
\begin{align}
 	\label{eq:PiReQuasiP}
	&\Pi^R_{\mathrm{e}}(t,p^{0},\vec p) 
  	=-\frac{1}{2\pi}\sum_{n=1}^{\infty}
  	\int 
	\prod_{i=1}^{2n}\frac{\dd{\mathbf{q}_{i}}}{(2\pi)^{d}2\omega(\mathbf{q}_{i})} \,
	\sum_{m=0}^{2n}
	\frac{\eta^{2n}}{m!(2n-m)!}
	\nonumber\\
	&\quad\times\
	\left(p^{0}-\omega_{\mathbf{q}_{1}}-\dots-\omega_{\mathbf{q}_{m}}
	+\omega_{\mathbf{q}_{m+1}}+\dots+\omega_{\mathbf{q}_{2n}}+\mathrm{i}\epsilon\right)^{-1}
	\nonumber\\
	&\quad\times\
	\delta(\mathbf{p}-\mathbf{q}_{1}-\dots-\mathbf{q}_{m}+\mathbf{q}_{m+1}+\dots+\mathbf{q}_{2n})
	\nonumber\\
	&\quad\times\
	\left[(f_{\mathbf{q}_{1}}+1)\cdots(f_{\mathbf{q}_{m}}+1)f_{\mathbf{q}_{m+1}}\cdots f_{\mathbf{q}_{2n}}
	\right.
	\nonumber\\
	&\quad\quad-\
	\left.
	f_{\mathbf{q}_{1}}\cdots f_{\mathbf{q}_{m}}(f_{\mathbf{q}_{m+1}}+1)\cdots 
	(f_{\mathbf{q}_{2n}}+1)\right]
	\,,
\end{align}
\begin{align}
 	\label{eq:PiRoQuasiP}
  	&\Pi^R_{\mathrm{o}}(t,p^{0},\vec p) 
  	=-\frac{1}{2\pi}\sum_{n=1}^{\infty}
  	\int 
	\prod_{i=1}^{2n+1}\frac{\dd{\mathbf{q}_{i}}}{(2\pi)^{d}2\omega_{\mathbf{q}_{i}}} \,
	\sum_{m=0}^{2n+1}
	\frac{\eta^{2n+1}}{m!(2n-m+1)!}
	\nonumber\\
	&\quad\times\
	\left(p^{0}-\omega_{\mathbf{q}_{1}}-\dots-\omega_{\mathbf{q}_{m}}
	+\omega_{\mathbf{q}_{m+1}}+\dots+\omega_{\mathbf{q}_{2n+1}}+\mathrm{i}\epsilon\right)^{-1}
	\nonumber\\
	&\quad\times\
	\delta(\mathbf{p}-\mathbf{q}_{1}-\dots-\mathbf{q}_{m}+\mathbf{q}_{m+1}+\dots+\mathbf{q}_{2n+1})
	\nonumber\\
	&\quad\times\
	\left[(f_{\mathbf{q}_{1}}+1)\cdots(f_{\mathbf{q}_{m}}+1)f_{\mathbf{q}_{m+1}}\cdots f_{\mathbf{q}_{2n+1}}
	\right.
	\nonumber\\
	&\quad\quad-\
	\left.
	f_{\mathbf{q}_{1}}\cdots f_{\mathbf{q}_{m}}(f_{\mathbf{q}_{m+1}}+1)\cdots 
	(f_{\mathbf{q}_{2n+1}}+1)\right]
  	  \,,
\end{align}
\end{subequations} 
which, due to the non-perturbative $s$-channel re-summation, take a similar form as the scattering integral \eq{Cfmain2} itself.
Note, however, that the on-shell energy conservation is replaced by the energy denominator shifted into the complex plane, which, in addition, includes a principal value part.
Moreover, due to the sums over $m$, the expressions still include inelastic such as $m\leftrightarrow2n-m$ processes.

\section{\label{sec:scalingSG}Non-thermal fixed points of the sine-Gordon model}
Having set up our non-perturbative kinetic theory of transport in the sine-Gordon model, the subsequent step is to examine the resulting kinetic equation for possible universal dynamics, i.e., scaling evolution in space and time at a non-thermal fixed point.
For this, one presupposes that, in a late-time scaling limit, the distribution function obeys a scaling form
\cite{Orioli:2015dxa,Chantesana:2018qsb.PhysRevA.99.043620}
\begin{align}
	\label{eq:scalrel}
	f(t,\mathbf{p})=(t/t_0)^\alpha f_\mathrm{s}\left([t/t_0]^\beta \mathbf{p}\right)\,,
\end{align}
where $t_{0}$ is some reference time within the scaling interval. 
$f_\mathrm{s}$ is a universal scaling function depending on momentum only, which defines, together with the scaling exponents $\alpha$ and $\beta$, the universality class of the respective non-thermal fixed point.
The task will be to determine the scaling exponents $\alpha$ and $\beta$ of the space-time rescaling, as well as the universal properties of the scaling function $f_\mathrm{s}$ by means of a scaling analysis of the kinetic equation \eq{pre_kinetic} with scattering integral \eq{Cfmain2} and $T$-matrix \eq{Tmatrix}, where the coupling functions are defined in Eqs.~\eq{LambdaReopMain}, \eq{PiReQuasiP}, and \eq{PiRoQuasiP}.
We will argue that there are different parameter regimes depending, in particular, on the total quasiparticle density and the renormal\bae{ise}{ize}d mass $M$, where different types of universal dynamics prevails.
As a consequence, different sets of scaling exponents will result, which hints to the existence of different non-thermal fixed points.

\subsection{Infrared fixed points: regimes of different scaling}
While, in general, the scaling evolution can be character\bae{ise}{ize}d by either a positive or a negative exponent $\beta$, our focus is set on infrared fixed points, i.e., evolutions towards low wave numbers, with $\beta>0$.
For momenta on the order of the mass, $p\approx M$, the dispersion $\omega(\mathbf{p})$ is not a homogeneous function of $p$.
Hence, scaling is expected to require either $p\ll M$ or $p\gg M$. 
At infrared fixed points, we can restrict ourselves to the non-relativistic limit $p\ll M$, where
\begin{align}
	\label{eq:nonreldisp}
	\omega(\mathbf{p})
	\approx M+\frac{\mathbf{p}^2}{2M}
	\,
\end{align}
throughout the range of momenta contributing to the scattering integral. 
In this limit, the energy-conservation delta distribution reduces scattering to on-energy-shell processes to which we have already restricted the integral in \Eq{Cfmain2}, and we can thus replace the distribution by
$\delta(\varepsilon_{\mathbf{p}}-\varepsilon_{\mathbf{q}_{1}}-\dots-\varepsilon_{\mathbf{q}_{n+1}}
	+\varepsilon_{\mathbf{q}_{n+2}}+\dots+\varepsilon_{\mathbf{q}_{2n+1}})$,
i.e., expressed in terms of the homogeneous non-relativistic dispersion 
\begin{align}
	\label{eq:scalrelDisp}
	\varepsilon(\mathbf{p})
	={\mathbf{p}^2}/{2M}
	=s^{-z} \varepsilon\left(s\, \mathbf{p}\right)
	\,,
\end{align}
with dynamic exponent $z=2$.
In contrast, the factors $1/2\omega(\mathbf{q}_i)$ in \eq{Tmatrix}, in the scaling limit, are approximately given by $1/2M$.
Analogously, the frequency-momentum integral over the statistical function, which enters, e.g., the dressed coupling \eq{lambdabarf},
\begin{align}
	\label{eq:Fxx}
	F(t,s=0) = F(x,x) 
	= \int\frac{\mathrm{d}^{d+1}p}{(2\pi)^{d+1}}F(t,p)
	\approx\frac{\eta n_{0}}{2M}\,,
\end{align}
is a constant which depends on the coupling, the dressed mass, and the quasiparticle density \eq{n0QP}.

Let us now have a look at the scattering integral \eq{Cfmain2} in order to identify a measure for distinguishing regimes which can give rise to different possible infrared fixed points.
The expression \eq{Cfmain2} contains an infinite sum over $(n+1)$-to-$(n+1)$ elastic collision processes between `particles', over all integer orders $n$.   
As we argue in the following, though, not all orders contribute with an equal weight. 
To estimate the relative importance of different orders, let us, for the first, assume that the order of magnitude of the coupling functions $\Lambda^R_{\mathrm{e/o}}$ is roughly equal for all $m$ and all $\sigma$. 
Under this assumption, the sum over the subsets $\sigma$ appearing in \eq{Tmatrix} is approximately proportional to the number of these subsets, such that their sum over $m$ in \eq{Tmatrix} evaluates to a power of $2$,
\begin{align}
	&\sum_{m=1}^{n}\left[\frac{(2n+1)!}{(2m)!(2[n-m]+1)!}+\frac{(2n+1)!}{(2m+1)!(2[n-m])!}\right]
	\nonumber\\
	&\quad 
	=2^{2n+1}-(2n+2) \approx 2^{2n+1}
	\,,\quad \text{for}\ \ n\gg1\,.
\end{align}
Estimating, furthermore, the momentum integrals over the distribution functions $f_{\mathbf{q}_{i}}$ to scale as the quasiparticle density $n_0$, cf.~\Eq{n0QP}, we find that the $n$th-order contribution to the scattering integral \eq{Cfmain2} scales as
\begin{align}
	C^{(n)}[f]
	\sim \left(\frac{\eta n_{0}}{M}\right)^{2n+1}\frac{1}{n!(n+1)!}
	\sim\frac{F_{0}}{n+1}\left(\frac{F_0^{n}}{n!}\right)^2
	\,,
	\label{eq:SGness}
\end{align}
where $F_0\equiv F(t,s=0)=\eta n_{0}/M$, recall \eq{n0QP}.
Hence, the dimensionless quantity $F_0$ provides a measure for the order $n$ of the interactions contributing to the cosine potential \eq{Vexpansion} of the sine-Gordon model, that dominates the collisional integral:  
If $F_0\ll 1$, all terms $C^{(n)}[f]$ beyond $n=1$ can be neglected, whereby we recover the standard wave-Boltzmann-type scattering integral of the $\phi^4$ theory, here describing elastic $2\to2$ scattering \cite{Orioli:2015dxa}. 

If instead $F_0\gg 1$, the sine-Gordon model gives rise to distinctly different collisional properties, because $C^{(n)}[f]$ reaches its maximum at $n\approx F_0\gg1$.
Consequently, the order $n$ of the dominant contribution can grow very large, in accordance with the exponential growth of the hyperbolic functions entering the collision integral \eq{Cfmain}. 
We will argue in the following that this can have a surprising effect on the scaling behavi\bae{our}{or} of the scattering integral as compared to the standard $\phi^4$ case.

\subsection{Spatio-temporal scaling analysis of the kinetic equation}
We now proceed to finding a scaling relation between the exponents $\alpha$ and $\beta$ by demanding both sides of the kinetic equation to show the same scaling.
Inserting the scaling form \eq{scalrel} into the kinetic equation \eq{pre_kinetic} and rescaling $\mathbf{p}\to(t/t_{0})^{-\beta}\mathbf{p}$ we obtain
\begin{align}
	\label{eq:ScalingKinEq}
	(t/t_{0})^{\alpha-1}\left[\alpha f_\mathrm{s}(\mathbf{p})+\beta\mathbf{p}\cdot\partial_{\mathbf{p}} f_\mathrm{s}(\mathbf{p})\right] 
	=t_{0}\,(t/t_{0})^{-\beta\mu} C[f_\mathrm{s}](\mathbf{p})
	\,.
\end{align}
Here, the exponent $\mu$ character\bae{ise}{ize}s the homogeneity of the scattering integral,
\begin{align}
	\label{eq:scalrelC}
	C[f](t,\mathbf{p})=(t/t_0)^{-\beta\mu} C[f_\mathrm{s}]\left([t/t_0]^\beta \mathbf{p}\right)
	\,.
\end{align}
It follows immediately from \Eq{ScalingKinEq} that a solution $f$ of the kinetic equation can assume the scaling form \eq{scalrel} if the scaling relation
\begin{flalign}
	\label{eq:scalcond}
	(\text{I})&&
	\alpha-1
	&=-\beta\mu
	&
\end{flalign}
between the three exponents is fulfilled.
This represents condition no.~I between the exponents $\alpha$ and $\beta$, with a yet to determine exponent $\mu$. 

To obtain a further relation between these exponents, one uses the explicit form \eq{Cfmain2} of the scattering integral for expressing $\mu$ in terms of the scaling exponents $\alpha$ and $\beta$ of $f$ and $z$ of $\omega(\mathbf{p})$.
In determining this relation, we proceed step by step, first having a look at a single contribution $C^{(n)}$ at order $n$ and focus on the gain and loss terms,
\begin{align}
	\label{eq:gainloss}
	&(f_{\mathbf{q}_{1}}+1)\cdots(f_{\mathbf{q}_{n+1}}+1)f_{\mathbf{q}_{n+2}}\cdots 
	f_{\mathbf{q}_{2n+1}}f_{\mathbf{p}}
	\nonumber\\
	&\quad-\ 
	f_{\mathbf{q}_{1}}\cdots f_{\mathbf{q}_{n+1}}(f_{\mathbf{q}_{n+2}}+1)\cdots 
	(f_{\mathbf{q}_{2n+1}}+1)(f_{\mathbf{p}}+1)
	\,.
\end{align}
At an infrared fixed point, we expect large occupation numbers, $f_\mathbf{q}\gg1$, throughout the region of low wave numbers $\mathbf{q}$.
In this limit, one can sort the different contributions to \eq{gainloss} with respect to powers in $f$. 
As usual, the highest-order terms, containing $2n+2$ distribution functions $f$ in the gain and loss terms cancel each other so that the leading-order terms consist of $2n+1$ $f$'s. 
Terms of lower order in $f$ can be neglected.

Note that in $2n+1$ out of these $2n+2$ terms of leading order $\mathcal{O}(f^{2n+1})$ in $C^{(n)}[f]$, i.e., in each of the terms in the sum in
\begin{align}
	\label{eq:LeadingOrderScattTerms}
	-\prod_{i=1}^{2n+1}f_{\mathbf{q}_{i}}
	+\sum_{j=1}^{2n+1}(-1)^{\theta(j-n-3/2)}f_{\mathbf{p}}\prod_{i=1\,|\,i\not=j}^{2n+1}f_{\mathbf{q}_{i}}
\end{align}
there is one `free' momentum $\mathbf{k}\equiv\mathbf{q}_{j}$, i.e., one momentum $\mathbf{k}\in\{\mathbf{q}_{1},\dots,\mathbf{q}_{2n+1}\}$ that is integrated over but which is not an argument of a distribution function $f(t,\mathbf{q}_{i})$.
Let us suppose that, if $n$ is sufficiently large, we can neglect, beyond these $2n+1$ terms, the single one without such a free momentum, i.e., the first term in \eq{LeadingOrderScattTerms}, which arises from the constant in the factor $f_\mathbf{p}+1$. 
Such a free momentum is already present in the collision integral of the standard $\phi^4$ theory, equivalent to the leading-order contribution $C^{(1)}[f]$. 
However, in the case of high `sine-Gordon-ness', $F_0\gg1$, cf.~our discussion following \Eq{SGness}, this free momentum can modify the scaling behavi\bae{our}{or} significantly.

The free momentum $\mathbf{k}$ is not constrained by the distribution $f$, neither by its IR scaling form, nor by its fall-off  in the UV. 
Nevertheless, in the standard $\phi^4$ collision integral, it does not cause a UV divergence, because it is still restricted by the delta distributions, which ensure momentum and energy conservation. 
In principle, this is true also for the full sine-Gordon integral. 
However, if the dominant order $n\sim F_0$ grows very large, these restrictions could \textit{effectively} disappear. 
The reason is that the large number of different momenta $\{\mathbf{p},\mathbf{q}_{i}\}$ and frequencies $\{\omega_{\mathbf{p}},\omega_{\mathbf{q}_{i}}\}$ can sum up, within the delta functions in \eqref{eq:Cfmain2}, to any, also very large total values $\mathbf{k}$ and $\omega_{\mathbf{k}}$, respectively. 
As a result, the single momentum $\mathbf{k}$ will then ultimately be restricted only by the UV cutoff of the theory. 
In that case, the integral over $\mathbf{k}$ does not contribute to the scaling of the scattering integral, other than in and close to the limit $n=1$.

Let us discuss this more rigorously. 
As introduced above, we denote free momentum by $\mathbf{k}$ while keeping the notation $\mathbf{q}_i$ for the other integrated ones and $\mathbf{p}$ for the external momentum. 
Neglecting the single term without a free momentum and pulling all integrals in $C^{(n)}$, except that over $\mathbf{k}$, as well as the $T$-matrix, the energy-momentum conservation and the gain and loss terms  into a kernel integral $K^{(n)}$, we rewrite the $n$-th order collision integral as
\begin{align}
	\label{eq:CfnofKfn}
	C^{(n)}[f](t,\mathbf{p})
	&\sim-\int \limits^{\Lambda_\text{UV}} d^dk\,
	K^{(n)}[f]\left(t,\mathbf{p};\varepsilon(\mathbf{k}),\mathbf{k}\right)
	f(t,{\mathbf{p}})
	\,,
\end{align}
with the kernel integral being defined as 
\begin{align}
	\label{eq:Kfn}
	&K^{(n)}[f](t,\mathbf{p};E,\mathbf{k})
	= 
	\int\prod_{i=1}^{2n+1}\frac{\dd{\mathbf{q}_{i}}}{(2\pi)^{d}} \,
	\sum_{m=1}^{2n+1}\Bigg\{
	\delta\left(\mathbf{k}-\mathbf{q}_{m}\right)
	\nonumber\\
	&\quad\times\
	\left|T^{(n)}(t;{\mathbf{p},\mathbf{q}_{1},\dots,\mathbf{q}_{2n+1}})\right|^{2}
	\nonumber\\
	&\quad\times\
	\delta\left(\mathbf{p}-\sigma^{(n)}_{m}\mathbf{k}-\sum\nolimits_{l=1;l\not=m}^{2n+1}\sigma^{(n)}_{l}\mathbf{q}_{l}\right)
	\nonumber\\
	&\quad\times\
	\delta\left(\varepsilon_{\mathbf{p}}-\sigma^{(n)}_{m}E
	          -\sum\nolimits_{l=1;l\not=m}^{2n+1}\sigma^{(n)}_{l}\varepsilon_{\mathbf{q}_{l}}\right)
	\nonumber\\
	&\quad\times\
	\sigma^{(n)}_{m}\prod_{l=1}^{m-1}f(t,{\mathbf{q}_{l}})\prod_{l'=m+1}^{2n+1}f(t,{\mathbf{q}_{l'}})
	\Bigg\}
	\,.
\end{align}
Here, the sign index $\sigma^{(n)}_{l}=1$ for $l=1,\dots,n+1$, and $\sigma^{(n)}_{l}=-1$ for $l=n+2,\dots,2n+1$, cares for the signs of the momenta and energies as well as of the gain vs. the loss terms.
Recall again that the free momentum $\mathbf{k}=\mathbf{q}_{j}$ is defined as the one for which, in each of the terms in the sum over $m$ the factor $f(t,\mathbf{k})$ is missing, cf.~\eq{LeadingOrderScattTerms}.

Assume, for a moment, that the $T$-matrix is a momentum-independent constant.
Since the distributions $f$ appearing in $K^{(n)}$ are the same for each integrated momentum $\mathbf{q}_{i}\not=\mathbf{k}$,  it becomes very reasonable, for $n\gg 1$, to adopt a statistical approach to the analysis of the highly intricate scattering integral.
We can thus apply the central-limit theorem and conclude that both $E$ and $\mathbf{k}$ will be distributed according to a multivariate, uncorrelated Gaussian, i.e., that the kernel distribution can be approximated by  
\begin{align}
	\label{eq:KfnGauss}
	&K^{(n)}[f](t,\mathbf{p};E,\mathbf{k})
	\sim  \exp\left(-\frac{E^2}{2\sigma_E^2}\right)\exp\left(-\frac{\mathbf{k}^2}{2\sigma_p^2}\right)
	\,,
\end{align}
where we neglected constants and where the standard deviations scale as $\sigma_E\sim [M+\varepsilon(p_{\Lambda})]\sqrt{n}$ and $\sigma_p\sim p_{\Lambda}\sqrt{n}$, with $p_{\Lambda}$ being a scale which character\bae{ise}{ize}s the width $\langle p^{2}\rangle_{f}=\int_\mathbf{p}p^{2}f(t,\mathbf{p})/\int_\mathbf{p}f(t,\mathbf{p})$ of the distribution $f(t,{\mathbf{p}})$ in momentum space. 
In accordance with the previous discussion, we assume this scale to be $p_{\Lambda}^{2}\ll M$.
Note, in advance, that, for the case of the spatial scaling form $f(t,{\mathbf{p}})\sim[p_{\Lambda}^{\kappa}+p^{\kappa}]^{-1}$ found in our numerical simulations \cite{Heinen:2022rew},
the width $\langle p^{2}\rangle_{f} \approx p_{\Lambda}^{2}$ is set by the infrared cutoff scale, below which the distribution is constant in $p$.

Consider a UV cutoff scale $\Lambda_\text{UV}$ that is set by the inverse spatial resolution with which one describes the system, e.g., a scale on the order of $M$, beyond which our low-energy effective theory analysis is expected to be inapplicable.
Then, if $\varepsilon(\Lambda_\text{UV})\ll \sigma_E$ and ${\Lambda}_\text{UV}\ll \sigma_k$, which will be the case for sufficiently large $n\approx F_{0}\gg(\Lambda_{\mathrm{UV}}/p_{\Lambda})^{2}$, the collision integral term \eq{CfnofKfn} reduces to
\begin{align}
	\label{eq:CfnofKfnLargenScaling}
	C^{(n)}[f](t,\mathbf{p})
	&\sim-\Lambda_\text{UV}^d\,
	K^{(n)}[f]\left(t,\mathbf{p};0,0\right)
	f(t,{\mathbf{p}})
	\,.
\end{align}

It remains to take into account that the $T$-matrix is, in general, not independent of its arguments. 
According to \Eq{Tmatrix}, it contains a sum of coupling functions $|\Lambda^R_\mathrm{e/o}|^{2}$ with different arguments. 
These arguments are sums of subsets of the arguments of the $T$-matrix, i.e., of $\mathbf{p}$, $\mathbf{k}$ and the $\mathbf{q}_i$, as well as of the respective energies.
Similarly to the above discussion for the kernel integral $K^{(n)}$, we may conclude that these arguments, according to the loop functions  $\Pi^R_{\mathrm{e/o}}$, Eqs.~\eq{PiReQuasiP}, \eq{PiRoQuasiP}, defining the coupling functions $\Lambda^{R}_{\mathrm{e/o}}$, \Eq{LambdaReopMain}, are also distributed, for sufficiently large $n$, as uncorrelated Gaussians. 
For sufficiently large $n\sim F_{0}$, their standard deviation will be much larger than ${\Lambda}_\text{UV}$, such that we can shift the energy and momentum arguments by $\varepsilon(\mathbf{k})$ and $\mathbf{k}$, respectively, without changing the integrals, such that we obtain that 
\begin{align}
	\left|T^{(n)}(t;{\mathbf{p},\mathbf{q}_{1},\dots,\mathbf{q}_{2n+1}})\right|_{\mathbf{q}_{m}=\mathbf{k}}
	\simeq\left|T^{(n)}(t;{\mathbf{p},\mathbf{q}_{1},\dots,\mathbf{q}_{2n+1}})\right|_{\mathbf{q}_{m}=0}
	\,,
\end{align}
for all $m=1,\dots,2n+1$.
Thereby the $T$-matrix becomes independent of $\mathbf{k}$ and the above analysis leading to the approximation \eq{CfnofKfnLargenScaling} remains valid.

The resulting approximate dependence \eq{CfnofKfnLargenScaling} of the collisional integral on the distribution function $f$, together with the full definition \eq{Kfn} of the kernel function enables us, finally, to derive its scaling exponent $\mu$ as defined in \Eq{scalrelC}.
Collecting all contributions from the remaining $2n$ momentum distribution functions and $\mathbf{q}_{i}$-integrals, from the $d$-dimensional momentum and $z$-dimensional energy delta functions, from the $T$-matrix, and from the outer $f(t,\mathbf{p})$, we find the exponent to be given by
\begin{align}
\label{eq:betamu}
	\mu
	&=2n(d-\alpha/\beta)-d-z+2m-\alpha/\beta
	\,,
\end{align}
where $m$ denotes the scaling exponent of the $T$-matrix,
\begin{align}
	\label{eq:Tscaling}
	&|T^{(n)}(t;\mathbf{p},\{\mathbf{q}_{i}\})|
	= s^{-m} |T^{(n)}(s^{-1/\beta}t;s\mathbf{p},\{s\mathbf{q}_{i}\})|	\,,
\end{align}
not to be confused with the sum indices $m$ used before.

Note that, upfront, we assume the exponent $m$ to be independent of $n$.
$\mu$, however, as given in \Eq{betamu}, is in general not independent of $n$, such that the entire collision integral $C[f]$, summed over all $n$ would not scale homogeneously. 
One typically postulates, though, that in the IR scaling region, either the total quasiparticle density \eq{n0QP}, or the quasiparticle energy density 
$\mathcal{E}_{0} = \int_\mathbf{p}\varepsilon(\mathbf{p})f(t,\mathbf{p})$
remains constant in time.
As, for the massive sine-Gordon model in the IR regime, where $\varepsilon(\mathbf{p})\ll M$, the collision integral is dominated by elastic, number conserving, i.e., $n$-to-$n$ processes, quasiparticle number \eq{n0QP} is conserved, $n_{0}\equiv\,$const., and thus the scaling form \eq{scalrel} is subject to the constraint
\begin{flalign}
	\label{eq:alphadbeta}
	(\text{II})&&
	\alpha
	&=d\beta
	\,. &
\end{flalign}

Therefore, the $n$-dependent term in \eq{betamu} drops out and $\mu$ results as being independent of $n$,
\begin{flalign}
\label{eq:betamuNumberconservation}
	(\text{III})&&
	\mu
	&=2m-d-z-\alpha/\beta
	\,.&
\end{flalign}

Interestingly, in contrast to the case of standard $\phi^{4}$ theory, one can also reverse this argument: 
The scaling hypothesis for the distribution function and thus for the scattering integral in fact requires that \eq{alphadbeta} holds and thus quasiparticle number \emph{must} be conserved. 
This is a remarkable feature in comparison with standard $\phi^4$ theory, where one has to \textit{postulate} particle conservation in the relevant momentum region \cite{Orioli:2015dxa}, i.e., to effectively restrict the collision integral to elastic $2$-to-$2$ processes.
In contrast to this, the transcendental dependence of the collision integral \eq{Cfmain} on $G^{\gtrless}$ shows already that for $C[f]$ to scale requires $G^{\gtrless}(x)$ to scale trivially, $G^{\gtrless}(t;x^{0},\mathbf{x})=G^{\gtrless}(s^{-1/\beta}t;x^{0}/s^{z},\mathbf{x}/s)$ (neglecting the fast oscillations with frequency $\sim M$), because otherwise, the hyperbolic functions prevent the collision integral to be homogeneous.
This trivial scaling, in turn, with Eqs.~\eq{Ggtrless}, \eq{f_def}, and the general scaling behavi\bae{our}{or} $\rho(p^{0},\mathbf{p})=s^{2}\rho(s^{z}p^{0},s\mathbf{p})$ of the spectral function, already leads to the scaling relation \eq{alphadbeta} and thus implies number conservation. 
We may conclude, therefore, that also for other theories, which lead to a collision integral depending transcendentally on $G^{\gtrless}$, or at least polynomially with a minimum of two different powers, scaling requires number conservation at a non-thermal fixed point. 
Vice versa, in the case that alternative conservation laws should hold, e.g., for the total energy, the collision integral can only depend on functions, whose arguments only involve the respective conserved `charge'.

We finally anal\bae{yse}{yze} the scaling of $\Pi^R_\mathrm{e/o}$, Eqs.~\eq{PiReQuasiP}, \eq{PiRoQuasiP}, in a similar manner. 
Again, a single integral over a `free' momentum does not contribute to the scaling if $F_0$ is sufficiently large.
As in the numerator of the scattering integral, all terms in the sum over $n$ scale in the same way by virtue of particle conservation. 
Finally, the scaling behavi\bae{our}{or} of the entire $T$-matrix is ensured by the fact that, for a high occupation $f(t,\mathbf{p})\gg1$ in the IR, close to an NTFP, we can approximate 
\begin{align}
	\bar{\lambda}\Pi^R_\mathrm{e/o}\gg 1\,,
\end{align}
such that $\Lambda^R_\mathrm{e/o}$ scales homogeneously if $\Pi^R_\mathrm{e/o}$ does so, cf.~\Eq{LambdaReopMain}. 
As a result, the energy denominator and the momentum conservation delta distribution in Eqs.~\eq{PiReQuasiP}, \eq{PiRoQuasiP} yield the scaling exponent of the $T$ matrix,
\begin{flalign}
	(\text{IV})&&
	m&=d+z\,.&
\end{flalign}
Together with this, we obtain, from conditions I-III, i.e.~Eqs.~\eq{scalcond}, \eq{alphadbeta}, and \eq{betamuNumberconservation}, the (anomalously small, see below) exponents $\alpha$ and $\beta$,
\begin{flalign}
	\text{(A)}&&
	\beta=\frac{1}{z+d}=\frac{1}{2+d}
	\,,
	&&
	\alpha=\frac{d}{z+d}
	\,.&&
	\label{eq:alphabetaanomalous}
\end{flalign}

We compare these with the `Gaussian' exponents, which we expect to be valid for small $F_{0}$ and thus $n\sim1$ dominating the collision integral.
In this case, the scaling of $C^{(n)}[f]$, under the assumption of number conservation, is found to be set by 
\begin{flalign}
\label{eq:betamuG}
	(\text{III-G})&&
	\mu_\mathrm{G}
	&=d+2n(d-\alpha/\beta)-d-z+2m-\alpha/\beta
	&
	\nonumber\\
	&&
	&=2m-z-\alpha/\beta
	\,,
\end{flalign}
since one more $\mathbf{q}_{i}$ integral, the one which dropped out in the anomalous case above, contributes at each order $n$ of the integral \eq{Cfmain2}.
Analogously, the scaling of the $T$-matrix is given by
\begin{flalign}
	(\text{IV-G})&&
	m_\mathrm{G}&=z\,,&
\end{flalign}
such that the  exponents $\alpha$ and $\beta$ for this `Gaussian' non-thermal fixed point \cite{Karl2017b.NJP19.093014,Chantesana:2018qsb.PhysRevA.99.043620,Mikheev:2018adp} result as
\begin{flalign}
	\text{(G)}&&
	\beta_\mathrm{G}=\frac{1}{z}=\frac{1}{2}
	\,,
	&&
	\alpha_\mathrm{G}=\frac{d}{z}
	\,.
	&&
	\label{eq:alphabetagaussian}
\end{flalign}
%

\subsection{Spatial scaling form}
Having determined the scaling exponents $\alpha$ and $\beta$ for the spatio-temporal universal scaling evolution of the occupation-number distribution $f(\mathbf{p})$ at an IR fixed point of the sine-Gordon model, it remains to find the scaling function $f_\mathrm{s}(\mathbf{p})$, cf.~\Eq{scalrel}.
In the following we anal\bae{yse}{yze} the kinetic equation at a fixed time for pure power-law solutions of the form 
\begin{align}
	\label{eq:powerlaw}
	f_\mathrm{s}(\mathbf{p})\sim |\mathbf{p}|^{-\kappa}\,.
\end{align}
Such a pure power law is typically observed only above a certain momentum scale,  $|\mathbf{p}|\gg p_\Lambda(t)\sim t^{-\beta}$, whereas $f(t,\mathbf{p})$ saturates in a plateau below this scale. 
In fact, as typically $\kappa\ge d$, the $\Pi^R_\mathrm{e/o}$ functions as well as the whole scattering integral would be IR divergent if a pure power law persisted to arbitrarily small momenta.
Instead, the plateau renders the integrals convergent in the IR, such that the scale $p_{\Lambda}$ takes the role of an IR regulator.
Typically, the scaling function therefore takes a form such as
\begin{align}
	\label{eq:fMK}
	f_\mathrm{s}(\mathbf{p})\sim \frac{p_{\Lambda}^{\kappa-d}}{p_{\Lambda}^{\kappa}+|\mathbf{p}|^{\kappa}}
	\,,
\end{align}
which needs to be normal\bae{ise}{ize}d such that it fulfils the condition \eq{n0QP}.
The way it goes over from the plateau into the power-law fall-off \eq{powerlaw} can be different, though \cite{Schmied:2018upn.PhysRevLett.122.170404}.

At the non-thermal fixed point, i.e., in the scaling limit $p_\Lambda\to0$ and thus $t\to\infty$, or, for all practical purposes, within the power-law region $|\mathbf{p}|\gg p_\Lambda$, we can make the ansatz \eq{powerlaw} and otherwise work with an IR momentum cutoff.
Similarly as in the spatio-temporal scaling analysis, we state the purely spatial scaling hypothesis that, at a fixed time, the scattering integral fulfils
\begin{align}
\label{eq:scalrelCspatial}
C[f](t_{0},\mathbf{p})=s^{-\mu_\kappa} C[f]\left(t_{0},s \mathbf{p}\right)
\,.
\end{align}
The time-independent fixed-point equation
\begin{align}
	\label{eq:ScalingFPEq}
	\left(\alpha +\beta\mathbf{p}\cdot\partial_{\mathbf{p}}\right) f_\mathrm{s}(\mathbf{p})
	=t_{0}C[f_\mathrm{s}](\mathbf{p})
	\,
\end{align}
then demands, if $\kappa\not=d$, that 
\begin{flalign}
	\label{eq:spatialscaleeq}
	(\text{V})&&
	\kappa
	&=-\mu_\kappa\,.
	&
\end{flalign}

Under a rescaling of all momenta, $\mathbf{p}\to s\mathbf{p}$, in the scattering integral \eq{Cfmain2}, the integral measures and the $f$'s, as well as in the $T$-matrix and the delta distributions, the $n$th-order term rescales according to 
\begin{align}
	\label{eq:prelimmukappa} 
        C^{(n)}[f](t_{0},\mathbf{p})
        &=s^{-2nd-2m_\kappa+d+z+(2n+1)\kappa} C^{(n)}[f]\left(t_{0},s \mathbf{p}\right)
	\,,
\end{align}
where we have again taken into account that the integral over the free momentum does not contribute to the scaling. 
In analogy to before, we denote the scaling exponent of the $T$-matrix at fixed time by $m_\kappa$, i.e.,
\begin{align}
	\label{eq:Tscalingspatial}
	&|T^{(n)}(\mathbf{p},\{\mathbf{q}_{i}\})|
	= s^{-m_\kappa} |T^{(n)}(s\mathbf{p},\{s\mathbf{q}_{i}\})|	\,.
\end{align}
However, \eq{prelimmukappa} cannot not be the final answer, yet, already because it is $n$-dependent. 
Besides this, we need to recall that the scattering integral in our case is non-local, i.e., it depends crucially on the physics at the IR boundary of the scaling interval, and we anticipate that the value of $C^{(n)}[f]$ at the fixed point is dominated by the values of $f$ and $T^{(n)}$ at the IR cutoff scale $p_{\Lambda}$. 
Hence, a rescaling of the momenta in the power-law region also includes a rescaling of the IR cutoff scale,
\begin{align}
	\label{eq:prelimmukappa2} 
        C^{(n)}[f](t_{0},\mathbf{p};p_{\Lambda})
        &=s^{-2nd-2m_\kappa+d+z+(2n+1)\kappa} C^{(n)}[f]\left(t_{0},s \mathbf{p};sp_{\Lambda}\right)
	\,.
\end{align}
To determine the actual scaling exponent $\mu_{\kappa}$ of $C^{(n)}[f]$, the scale $p_{\Lambda}$ must be held fixed, though, i.e., we need to determine the rescaling of the scattering integral under $p_{\Lambda}$ alone.
This can be inferred from the degree of IR divergence when replacing all $f$'s by a pure power law \eq{fMK} and imposing $p_{\Lambda}$ as an IR integral cutoff,
\begin{align}
	\label{eq:CnIRdivergence} 
        C^{(n)}[f](t_{0},\mathbf{p};p_{\Lambda})
        &=s^{-(2n-1)(d-\kappa)} C^{(n)}[f]\left(t_{0},\mathbf{p};sp_{\Lambda}\right)
	\,.
\end{align}
Here we took into account that, in any of the terms, if $\kappa>d$, there are at most $2n-1$ algebraically divergent integrals over functions $f\sim p^{-\kappa}$, each contributing a leading-order dependence $\sim p_{\Lambda}^{d-\kappa}$ on the cutoff.

To understand this argument, note that,
(\textit{i}), in each of the $2n+1$ out of $2n+2$ leading-order terms in the gain-minus-loss factor \eq{gainloss} that contain an $f(\mathbf{p})$ evaluated at the fixed momentum $\mathbf{p}$, there are $2n$ factors $f_{\mathbf{q}_{i}}$ being integrated over, which can, in principle, give a divergence.
And, (\textit{ii}), in the single term without an $f(\mathbf{p})$, there are $2n+1$ such factors. 
In any of the terms, however, not all of these integrals can diverge simultaneously:
In case (\textit{i}), the strongest divergence can be expected when $2n-1$ momentum integrals are dominated by the momentum regime $q_{i}\lesssim p_{\Lambda}$.
This is possible, if one of the two delta distributions constrains the free momentum $\mathbf{k}$, which is not weighted by an $f$, and the other one constrains one of the $f$-weighted momenta $\mathbf{q}_{j}$ to be of the order of magnitude of the outer momentum $|\mathbf{p}|\ge p_\Lambda$.
In the single term which represents case (\textit{ii}), there is no $f_{\mathbf{p}}$, such that the two delta distributions constrain two integrated momenta that are arguments of an $f$. 
Hence, in both cases, at most $2n-1$ of the $f$ functions can contribute to the overall IR-divergence.
 
In summary, if $\kappa> d$, the degree of algebraic divergence is $(2n-1)(\kappa-d)$. 
Subtracting this from the scaling exponent of $C^{(n)}$ in \eq{prelimmukappa2}, we obtain the $n$-independent exponent 
\begin{flalign}
	\label{eq:mukappa}
	(\text{VI})&&
	\mu_\kappa
	&=2m_\kappa-z-2\kappa
	\,.
	&
\end{flalign}

Note finally, that in principle, the scattering integral could also be dominated by the regime where only two of the momenta are larger than $p_{\Lambda}$, viz.~$\mathbf{p}$ and one further momentum $\mathbf{q}_{i}$. 
This would correspond to near-forward scattering where particles are scattered to modes close to the incoming momentum, with momentum transfer being $0<\Delta p\ll p_{\Lambda}$.
We will exclude this possibility, which is known to give rise to a wave-turbulent cascade, where the distribution $f_{\mathbf{p}}(t)$, in the regime $p\gg p_{\Lambda}$, remains by definition constant in time due to the local wave-turbulent transport. 
In that case, combining Eqs.~\eq{scalrel}, \eq{alphadbeta}, and \eq{fMK}, the spatial exponent must be $\kappa=d$ \cite{Zakharov1992a,Chantesana:2018qsb.PhysRevA.99.043620}, such that, in this case the divergence does not alter the scaling \eq{prelimmukappa2}.

In order to obtain a prediction for $\kappa$, we furthermore need to anal\bae{yse}{yze}, in the same way, the functions  $\Pi^R_\mathrm{e/o}$ and thus $T^{(n)}$ to find $m_{\kappa}$. 
As the argument is similar for $\Pi^R_\mathrm{e}$ \eq{PiReQuasiP} and $\Pi^R_\mathrm{o}$ \eq{PiRoQuasiP}, we describe it only for the former. 
Including any divergent integral but the free one, upon rescaling we find that $-m_{\kappa}={(2n-2)d-z-(2n-1)\kappa}$. 
We expect the degree of IR divergence, i.e., the scaling exponent of $p_{\Lambda}^{-1}$ to be $(2n-1)(\kappa-d)$ and thus the same as above, which ensures that the overall scattering integral remains finite in the limit $p_{\Lambda}\to0$. 
The crucial difference to the numerator of the scattering integral is that, on the one hand, there is no $f$ in $\Pi^R_\mathrm{e/o}$, the argument of which is an outer momentum. 
On the other hand, there is no energy conserving delta distribution that can prevent one integral to be dominate by $f$ being evaluated in the IR limit. 
One can still argue that the $\mathrm{i}\epsilon$-regulated energy denominator at least suppresses configurations with the sum of the energies being large and that therefore at least one momentum should be of the order of the outer momentum and thus not come close to $p_{\Lambda}$. 

Subtracting the contribution to scaling from $p_{\Lambda}$ alone, we again obtain an $n$-independent result,
\begin{flalign}
	(\text{VII})&&
	m_\kappa
	&=z+d\,.
	&
\end{flalign}
Inserting this into \Eq{mukappa} finally gives, with \Eq{spatialscaleeq}, the spatial exponent at the anomalous non-thermal fixed point,
\begin{flalign}
	\text{(A)}&&
	\kappa=2d+z=2d+2
	\,.&&
	\label{eq:kappaanomalous}
\end{flalign}
Again, this is in contrast with the Gaussian fixed point, which we expect to play a role for small $F_{0}$ and thus $n\sim1$ dominating the collision integral.
In that case, one finds the weaker fall-off with, cf.~Ref.~\cite{Chantesana:2018qsb.PhysRevA.99.043620},
\begin{flalign}
	\text{(G)}&&
	\kappa_\mathrm{G}=d+z/2=d+1
	\,.&&
\end{flalign}
We note that also in this Gaussian limit, the scaling at a fixed time $t_{0}$ is partially determined by the dependence on $p_{\Lambda}$, which needs to be removed in order to obtain $\kappa$.
Notably, in this Gaussian case, only the effective coupling, i.e., $\Pi^{R}$, is dominated by one integral momentum below $p_{\Lambda}$, while there are no relevant divergences in the scattering integral. 
This can be seen from the explicit calculation of the integral \cite{Chantesana:2018qsb.PhysRevA.99.043620}, which demonstrates that a down-conversion like one-to-two process where only one momentum in the elastic collision vanishes, while the two of the momenta add up to the third one, is possible but kinematically suppressed.
This constitutes a further qualitative difference to the scaling we here propose to characterise universal sine-Gordon  dynamics.

\subsection{Scaling analysis in position space and time}
\label{sec:PositionSpace}
We finally re-examine the above scaling analysis, thereby considering the gain and loss terms in the scattering integral \eq{Cfmain} in \emph{position space} and time.
In this way, we can exhibit more clearly the spatially local character of the scattering processes in the case of the anomalously slow scaling with exponents  \eq{alphabetaanomalous} and \eq{kappaanomalous}, which stands in contrast to wave-turbulent types of scaling evolution that are rather \emph{local in momentum space}. 

In deriving the scattering integral in the form \eq{CfnofKfn}, with kernel \eq{Kfn}, we expanded the gain and loss terms \eq{gainloss} to leading order in powers of the large occupation numbers $f$, involving $2n+1$ factors $f_{\mathbf{q}_{i}}$, i.e., one less than there are factors in each term in \eq{gainloss}.
This expansion, in fact, can already be done in the scattering integral \eq{Cfmain}, where it corresponds to keeping only terms where all but one factor of $\rho$ are paired with one factor of $f$:
Using \eq{Ggtrless}, we can replace $G^{<}$  by $G^{<}(t,p)= G^{>}(t,p)-\mathrm{i}\rho(p)=-\mathrm{i}\left[f(t,p)+1\right]\,\rho(t,p)$.
Expanding then the scattering integral to first order in the added $-\mathrm{i}\rho$, i.e., to order $\mathcal{O}(1/f)$, one obtains
\begin{align}
	\label{eq:Cfmainrho1}
	&C[f](t,\mathbf{p})
	=
	\mathrm{i}\int_0^{\infty} \frac{\dd{p^0}}{2\pi} \,
	\mathcal{F}\Bigg[
	\left\{\left[\left|\Lambda^{R}_{\mathrm{e}}\right|^{2}\ast\left(\rho\cdot\sinh G^{>}\right) \right]\cdot\sinh G^{>}\right.
	\nonumber\\
	&\qquad\quad+\
	\left[\left|\Lambda^{R}_{\mathrm{e}}\right|^{2}\ast\left(\cosh G^{>} - 1\right)\right]\cdot\rho\cdot\cosh G^{>}
	\nonumber\\
	&\qquad\quad+\
	\left[\left|\Lambda^{R}_{\mathrm{o}}\right|^{2}\ast\left(\rho\cdot\cosh G^{>} - \rho\right)\right]\cdot\cosh G^{>}
	\nonumber\\
	&\qquad\quad+\
	\left.\left[\left|\Lambda^{R}_{\mathrm{o}}\right|^{2}\ast\left(\sinh G^{>} - G^{>}\right)\right]\cdot\rho\cdot\sinh G^{>}\right\}
	\ast G^{>} 
	\nonumber\\
	&\qquad\quad\!\!
	+\
	\left\{\left[\left|\Lambda^{R}_{\mathrm{e}}\right|^{2}\ast\left(\cosh G^{>} - 1\right)\right]\cdot\sinh G^{>}\right.
	\nonumber\\
	&\qquad\quad+\
	\left.\left[\left|\Lambda^{R}_{\mathrm{o}}\right|^{2}\ast\left(\sinh G^{>} - G^{>}\right)\right]\cdot\cosh G^{>}\right\}
	\ast \rho\Bigg](t,p)
	\nonumber\\
	&\qquad\quad+\ \text{subleading terms} \sim \mathcal{O}(f^{-2})
	\,,
\end{align}
because the leading terms containing $G^{>}\sim f\rho$ only cancel exactly in \eq{Cfmain}.
Note that we have expanded all hyperbolic functions of $\rho$ to linear order in $\rho$.
Moreover, we have written all terms of the integrand as a Fourier transform of the respective terms expressed in space and time, meaning that, as compared with \eq{Cfmain}, convolutions $\ast$ are exchanged with products $\cdot$ everywhere, and vice versa.
The scattering integral, expanded in this form, does no longer distinguish between gain and loss terms.
It rather represents the contribution to their difference, which is linear in an unpaired $\rho$.
Otherwise only $G^{>}$ appears, in which $\rho$ is multiplied with $f$.

In the integrand of \eq{Cfmainrho1}, the two-point functions $G^{>}(t,x)$ and $\rho(x)$ depend on $x=(x_{0},\mathbf{x})$, i.e., (relative) time $x_{0}$ and spatial coordinate $\mathbf{x}$ \footnote{%
In Eqs.~\eq{FrhoXp}, we denoted this coordinate as $s=(s_{0},\mathbf{s})$ but rename it here to not confuse it with the scaling parameter $s$ appearing below.}.
This has the advantage that the hyperbolic functions can be evaluated without convolutions in a Taylor expansion in powers of $G^{>}$. 
From the momentum-space expressions of $G^{>}$, \eq{Ggtrless}, and of the free spectral function, \eq{rho_free}, one finds
\begin{subequations}
	\label{eq:Ggtrofsrhoofs}
\begin{align}
	\label{eq:Ggtrofs}
	G^{>}(t,x) 
	&= \eta\int_\mathbf{p} \frac{f(t;\omega_{\mathbf{p}},\mathbf{p})}{\omega_{\mathbf{p}}}\,
	     \cos(\omega_{\mathbf{p}}x_{0}-\mathbf{p}\cdot\mathbf{x})
	\,,
	\\
	\label{eq:rhoofs}
	\rho(x) 
	&= \eta\int_\mathbf{p} \frac{1}{\omega_{\mathbf{p}}}\,
	     \sin(\omega_{\mathbf{p}}x_{0}-\mathbf{p}\cdot\mathbf{x})
	\,.
\end{align}
\end{subequations}
These expressions help exhibiting the \emph{local character, in position space},  of the scattering integral.
Both functions show fast temporal oscillations in $x_{0}$ with frequency $\simeq\omega_{0}=M$, which are modulated by the contributions from non-zero $\mathbf{p}$.

Consider, in particular, the case of anomalous scaling (A).
If $f$ has a universal form similar to \eq{fMK}, with $\kappa=2(d+1)$, \eq{kappaanomalous}, the function \eq{Ggtrofs} is strongly peaked at $x=0$, falling to small values on a scale set by $1/p_{\Lambda}$ in $|\mathbf{x}|$ and $2M/p_{\Lambda}^{2}$ in $x_{0}$.
In fact, one may easily show that $G^{>}(t;0,\vec x)\simeq F_{0}\exp\big[-(p_{\Lambda}|\vec x|)^{2}/2d\big]$,  at short distances $|\vec x|\lesssim 4p_{\Lambda}^{-1}$, and $G^{>}(t;x_{0},0)\simeq F_{0}\cos\big[(M+p_{\Lambda}^{2}/2M)x_{0}\big](\cosh[p_{\Lambda}^{2}x_{0}/2Mc_{d}])^{-1}$, with $F_{0}=\eta n_{0}M^{-1}$ and some $\mathcal{O}(1)$ constant $c_{d}$ for $x_{0}\lesssim 4M/p_{\Lambda}^{2}$.

Recall now that, as we found above, the anomalous scaling with \eq{alphabetaanomalous} and \eq{kappaanomalous} requires $F_0 \gg 1$.
Hence, the hyperbolic functions in \eq{Cfmainrho1} are exponentially large at the peak, where $G^{>}(t,0)\simeq F(t,0)=F_{0}$, while the spectral function $\rho$ outside the exponentials remains of order $\mathcal{O}(1)$. 
Moreover, the hyperbolic functions are approximately equal, $\cosh G^{>}\simeq\sinh G^{>}\simeq\exp(G^{>})/2$ and thus fall off to exponentially smaller values at non-zero $x$, such that one may neglect all contributions beyond the above scales. 
In fact, making use of the regular\bae{is}{iz}ed representation $\delta_{\epsilon}(x)=(2\pi\epsilon)^{-1/2}\exp(-x^{2}/2\epsilon)$ of the Dirac distribution, $\delta(x)=\lim_{\epsilon\to0}\delta_{\epsilon}(x)$, we can approximate the hyperbolic functions, in the limit $F_{0}\gg1$, as 
\begin{align}
  \exp[G^{>}(t,x)]
  \approx2\pi\sqrt{\epsilon_{0}\epsilon}\,\mathrm{e}^{F_{0}}C(x_{0})\,\delta_{\epsilon_{0}}(x_{0})\,\delta_{\epsilon}(|\mathbf{x}|)
  \,, 
\label{eq:expGgrtDirac}
\end{align}
with $\epsilon_{0}=(2Mc_{d})^{2}/(p_{\Lambda}^{4}F_{0})$, $\epsilon=d/(p_{\Lambda}^{2}F_{0})$, where $c_{d}$ is a $d$-dependent constant.
We have, in particular, separated off the fast oscillations with frequency $\omega_{\Lambda}=M+p_{\Lambda}^{2}/2M\gg p_{\Lambda}^{2}/2M$ into the function 
\begin{align}
   C(x_{0})
   =\sum_{n\in\mathbb{Z}}\sqrt{\gamma}\,\delta_{\gamma}\left(x_{0}-\frac{2\pi n}{\omega_{\Lambda}}\right)\,
   w_{\pi/\omega_{\Lambda}}\left(x_{0}-\frac{2\pi n}{\omega_{\Lambda}}\right)
   \,, 
\end{align}
with window function $w_{\delta}(x)=\theta(x+\delta)\theta(\delta-x)$.
In the limit $F_{0}\gg1$, $M\gg p_{\Lambda}$, $C(x_{0})$ is a Dirac comb with $\gamma=2\omega_{\Lambda}^{-2}/F_{0}$.
Hence, notwithstanding these fast oscillations, as a result, the hyperbolic functions in the scattering integral are, to a good approximation, local in $x$, such that the terms in curly brackets in \eq{Cfmainrho1} reduce to constants multiplying the functions $G^{>}$ and $\rho$, respectively.
These constants depend on $F_{0}$ and $p_{\Lambda}$, respectively, and the degree of spatial locality is set by $\epsilon_{0}$ and $\epsilon$, and thus the stronger, the larger $F_{0}$ is.
This \emph{spatio-temporal locality} is in stark contrast to the structure of scattering integrals giving rise to wave-turbulent transport, in which typically forward scattering dominates and which thus are rather local in \emph{momentum} space.
 
To analyse and discuss quantitatively the consequences of this locality is beyond the scope of the present work.  
The above analysis, however, also enables us to shed more light on the origins of the anomalous scaling, as we will discuss in the remainder of this section.

In principle, the constant rest mass $M$ prevents an exact scaling relation of the above functions and thus of the scattering integral.
In our previous scaling analysis, we neglected this, however, by considering the non-relativistic limit where only  on-mass-shell collision processes play a role in the dynamics, such that an equal number of positive and negative frequencies appears in the energy conservation delta distribution in \eq{Cfmain2}. 
This means, that only those terms contribute to \eq{Cfmainrho1}, in which the fast oscillations with frequency $M$ interfere with each other destructively.
So, in order to do obtain the scaling properties of \eq{Cfmainrho1}, we need to disregard the fast rest-mass contribution to the expanded frequency \eq{nonreldisp} in the arguments of the trigonometric functions in Eqs.~\eq{Ggtrofsrhoofs}.
In the frequency $\omega_{\mathbf{k}}$ in the denominators in \eq{Ggtrofsrhoofs}, however, we neglect the kinetic energy $p^{2}/2M$ as compared with the much larger $M$.
Under these approximations, the functions scale as
\begin{subequations}
\begin{align}
	\label{eq:GgtrofsScaling}
	G^{>}(t,x_{0},\mathbf{x}) 
	&= s^{d-\alpha/\beta}G^{>}(s^{\beta}t,x_{0}/s^{z},\mathbf{x}/s)
	\,,\\
	\label{eq:rhoofsScaling}
	\rho(t,x_{0},\mathbf{x}) 
	&= s^{d}\rho(s^{\beta}t,x_{0}/s^{z},\mathbf{x}/s)
	\,.
\end{align}
\end{subequations}
Since $G^{>}$ appears, in \eq{Cfmainrho1}, as an argument in the hypergeometric functions, it must not rescale with $s$, which immediately returns the condition \eq{alphadbeta}, that $\alpha/\beta=d$.

In contrast, the spectral function $\rho$ scales with $s^{d}$ and gives rise to an overall contribution of the dimension $d$ to the homogeneity exponent $\mu$ of the scattering integral, cf.~\eq{scalrelC}.
It is this contribution which gives rise to the distinction between the Gaussian, $\beta=1/z$, and the anomalous exponent, $\beta=1/(z+d)$, derived before.

Let us have a closer look at the relevance of $\rho$ in \eq{Cfmainrho1}.
Since $\rho$ is not restricted by the distribution $f$, higher momenta contribute to it more significantly than to $G^{>}$. 
Hence, $\rho(x)$ falls off to smaller values and oscillates on even shorter time and length scales, approximately given by the UV cutoff beyond which the low-energy-effective description is not expected to apply.
This cutoff, in our case, is estimated to be $\Lambda_\mathrm{UV}\lesssim M$. 

$\rho$, however, in the scattering integral \eq{Cfmainrho1}, is not included in the argument of the exponentially large hyperbolic functions, but rather multiplies these.
Hence, compared with the amplitude of the exponentiated $G^{>}$, cf.~\Eq{expGgrtDirac}, $\rho$ can be approximated as a constant,
\begin{align}
\label{eq:rhotimesexpGgtrofs}
	\rho(x)\exp\left[G^{>}(t,x)\right] 
	&\simeq2\rho^{+}(0)\exp\left[G^{>}(t,x)\right]
	\,,
\end{align}
where $\rho^{+}(0)=\eta\int_{\mathbf{p}}(2\omega_{\mathbf{p}})^{-1}$ is the positive-frequency contribution at $x=0$.
This approximation is equivalent to integrating over the free momentum $\mathbf{k}$ after using the central-limit theorem to arrive at \Eq{KfnGauss} \footnote{Taking only the positive-frequency, or, equivalently, the negative-frequency part into account is necessary since $\rho(0)=0$. Moreover, because energy conservation requires all collisions to be on mass shell, the fast oscillations with zero-momentum frequency $M$ are irrelevant for our subsequent scaling analysis.}.

As a result, in the case that $F_{0}\gg1$, the spectral function $\rho$ takes the role of a constant pre-factor in the scattering integral and thus, it does not rescale with $s^{d}$, cf.~\Eq{rhoofsScaling}, as it would be the case when $F_{0}\simeq1$ and thus the hyperbolic functions can be approximated to lowest order. 
Hence, in the anomalous case, the scaling exponent $\mu$ of $C[f](t,\mathbf{p})$ is diminished as described above, by the dimension $d$. 

We finally remark that a similar analysis applies to the coupling functions \eq{LambdaReopMain} appearing in \eq{Cfmainrho1}, as we can expand the re-summed `loop' functions in the same way and write them, in position space, as
\begin{subequations}
\begin{align}
  &\Pi^R_{\mathrm{e}}(t,p^{0},\vec p) 
  = \mathcal{F}\left[\theta\cdot\left(\rho\cdot\sinh G^{>}\right)\right](t,p^{0},\vec p) 
    \,,
    \\
  &\Pi^R_{\mathrm{o}}(t,p^{0},\vec p) 
  = \mathcal{F}\left[\theta\cdot\left(\rho\cdot\cosh G^{>}-\rho\right)\right](t,p) 
    \,.
    \label{eq:PiRdefrho1}
\end{align}
\end{subequations}
Again, the function $\rho$ can be approximated as a constant and its contribution to the scaling be dropped.

\section{\label{sec:Conclusions}Summary and Outlook}
We have extended the non-perturbative theory of universal scaling at infrared non-thermal fixed points to account for the non-polynomial interactions of the sine-Gordon model.
At an infrared fixed point, the system's excitations are character\bae{ise}{ize}d by large overoccupation of low-wave-number modes, that typically leads to a strong renormal\bae{isation}{ization} of the effective interactions governing the dynamics.
Within the 2PI effective action approach, we have chosen a re-summation scheme, which accounts for such a renormal\bae{isation}{ization}.
While such schemes are made and usually used for field theoretic models with simple, low-order polynomial interactions, we have achieved a closed-form re-summation that takes into account the non-polynomial potential of the sine-Gordon model.
The series re-summation thus entails not only loop diagrams containing self-consistently dressed propagators up to arbitrary orders in the coupling $\lambda$ but also re-sums the bare couplings themselves to arbitrary order.
We have used the resulting non-perturbative 2PI effective action to derive Kadanoff-Baym equations and from these a wave-Boltzmann-type kinetic equation.

The resulting wave-kinetic equation is reminiscent of those used for wave turbulence, both in the weak- and strong-coupling limits, and accounts for elastic collisional interactions of arbitrary many quasiparticle modes.
Using a standard scaling analysis we were able to determine the possible universal scaling solutions that represent a self-similar transport of excitations to lower wave-numbers.
For small quasiparticle densities, we could recapture Gaussian scaling that is character\bae{ise}{ize}d by the same exponents as for a standard $\phi^{4}$ model.
While, in this limit, the weakly varying field explores only the low-order Taylor contributions to the cosine potential, different scaling is predicted to prevail in the opposite limit, when the excitations probe many minima of the periodic interactions.
In this situation, energy and momentum conservation in the collisions are found to constrain the momenta of the participating modes only weakly.
This gives rise to a rather non-local transport in momentum space and an anomalously slowed scaling evolution.
Our analytic predictions made here are corroborated by numerical simulations reported in \cite{Heinen:2022rew}, which provide an intuitive picture of the coarsening dynamics underlying the anomalous non-thermal fixed point.

We remark that ultimately, a renormali\bae{s}{z}ation-group (RG) formulation of non-thermal fixed points, in analogy to fixed points describing, e.g., phase transitions in equilibrium, would be desirable.
While there has been some progress \cite{Gasenzer:2008zz,Berges:2008sr,Mathey2014a.PhysRevA.92.023635,Mikheev2023a}, such a formulation is lacking to date.
Note, also, that the $s$-channel re-summation of the 2PI effective action, for the case of a $\varphi^{4}$ theory, has been shown to correspond to solving coupled functional RG equations for the 2- and 4-vertices \cite{Gasenzer:2008zz}.
Moreover, numerical results confirm or support the existence of the predicted scaling solutions of the kinetic equations \cite{Walz:2017ffj.PhysRevD.97.116011,Chantesana:2018qsb.PhysRevA.99.043620}, and therefore, it is expected that also the scaling behaviour derived in the present work represents a leading-order result which will have to be refined by more advanced methods once available.

The results presented here suggest that the methods used could be applicable also to other models with transcendental interactions containing Taylor vertex terms to arbitrary order in powers of the field.
Moreover, together with the numerical simulations, they open a perspective on describing coarsening dynamics and phase-ordering kinetics from first principles.
This could also play a role for systems not directly associated with a model such as sine-Gordon.
For example, the sine-Gordon model examined here is known to serve, in equilibrium, as an effective model that captures phase transitions and excitations governed by ensembles of non-linear excitations such as vortices to which it is related by a duality transformation 
\cite{%
Berezinskii1971JETP...32..493B,
Berezinskii1972JETP...34..610B,
Kosterlitz1973a,
Jose1977a.PhysRevB.16.1217,
Minnhagen1987a.RevModPhys.59.1001,
Barmettler2010a,
Kosterlitz2020a.JLTP,
Nogueira2005a.PhysRevB.72.014541,
Nogueira2006a.PhysRevB.73.104515%
}.
Provided such transforms to apply (approximately) between models defined in Minkowskian space time, a direct application of the methods presented here appears possible. 
In fact, the exponents obtained in the present work for the case of two spatial dimensions ($\beta=1/4$, $\alpha=1/2$, and $\kappa=6$) are surprisingly close to those found at the ``strongly anomalous non-thermal fixed point" described in \cite{Karl2017b.NJP19.093014} for a two-dimensional Bose gas.
At that fixed point, the universal dynamics is found to be associated with strong clustering of vortices of equal winding number $\pm1$ and characteri\bae{s}{z}ed by $\beta=0.193\pm0.05$, $\alpha=0.402\pm0.05$, and $\kappa\equiv\zeta_\mathrm{a}=5.7\pm0.3$. 
While this could be a coincidence, it leaves open the possibility of a deeper connection between the well-known dualities relating  topological defects to field theories of the sine-Gordon type in equilibrium and some  (approximate) version of theirs out of equilibrium.

\begin{acknowledgments}
The authors thank S.~Bartha, J.~Berges, A.~Chatrchyan, L.~Chomaz, Y.~Deller, J.~Dreher, K.~Geier, P. Gro{\ss}e-Bley, M.~Karl, H.~K\"oper, S.~Lannig, I-K. Liu, M.~K.~Oberthaler, J.~M. Pawlowski, A.~Pi{\~n}eiro Orioli, M.~Pr\"ufer, N.~Rasch, C.~M.~Schmied, I.~Siovitz, H.~Strobel, and S.K.~Turitsyn for discussions and collaboration on related topics. 
They acknowledge support 
by the ERC Advanced Grant EntangleGen (Project-ID 694561), 
by the Deutsche Forschungsgemeinschaft (DFG, German Research Foundation), through 
SFB 1225 ISOQUANT (Project-ID 273811115), 
grant GA677/10-1, 
and under Germany's Excellence Strategy -- EXC 2181/1 -- 390900948 (the Heidelberg STRUCTURES Excellence Cluster), 
and by the state of Baden-W{\"u}rttemberg through bwHPC and the DFG through 
grant INST 35/1134-1 FUGG (MLS-WISO cluster) 
and grant INST 40/575-1 FUGG (JUSTUS 2 cluster).
A.~N.~M. acknowledges financial support by the IMPRS-QD (International Max Planck Research School for
Quantum Dynamics).\\
\end{acknowledgments}

\begin{appendix}
\begin{center}
\textbf{APPENDIX}
\end{center}
\renewcommand{\theequation}{A\arabic{equation}}
\setcounter{equation}{0}
\setcounter{table}{0}
\makeatletter
%

\section{Notation}
\label{app:notation}
Choosing the ($+$\,$-$\,$-$\,$-$) convention for the metric, 
the Minkowski product of ($d+1$)-vectors
$p = (p^{0},p^{1},...,p^{d}) = (p^{0},\vec p) = (\omega,\vec p)$, etc.
reads
$px = p^{0}x^{0} - \vec p\cdot\vec x$.
Defining the $(d+1)$-dimensional Fourier transform as 
$\mathcal{F}[f(p)](x) = f(x) = \int_{p}\exp\{-ipx\}f(p)\equiv (2\pi)^{-(d+1)}\int{\mathrm{d}^{d+1}p}\exp\{-ipx\}f(p)$,
the following convention is used for convolutions:
\begin{align}
  	(f\ast h)(x) 
  	&= \int\df{y}f(x-y)\,h(y)
	\label{eq:notConvolution}
  	\,,\\
  	(f\ast h)(p) 
  	&= \int\frac{\df{q}}{(2\pi)^{d+1}}f(p-q)\,h(q)
	\,.
\end{align}
The convolution theorem is then
\begin{align}
  	\mathcal{F}[(f\ast h)](x) 
  	&= (f\cdot h)(x) = f(x)\, h(x)
  	\,,\\
  	\mathcal{F}[(f\ast h)](p) 
  	&= (f\cdot h)(p) = f(p)\, h(p)
	\,.
\end{align}
For two-point functions in position space, we use the notation
\begin{align}
	\label{eq:notConvolutionTwoPoint}
  	(f\cdot h)(x,y) 
  	&=  f(x,y)\, h(x,y)
  	\,,\\
  	(f\ast h)(x,y) 
  	&= \int\df{z}f(x,z)\,h(z,y)
	\,,
\end{align}
which translates into the above convolutions in the case of translational invariance, $f(x,y)\equiv f(x-y)$.

\section{Decomposition of the self-energies}
\label{app:KBEq}
In this appendix, we provide details of the steps leading to the decomposition of the non-local self-energies according to \Eq{self_energy_decomp}, cf.~\Sect{KBEq}. 
Recall the decomposition   \eq{Pi}
of the loop functions \eq{LoopFcts}  into their spectral and statistical components using \eq{G_decomp} and standard trigonometric identities,
\begin{subequations}
	\label{eq:app:Pi}
	\begin{align}
		\label{eq:app:Pi_F_e}
		\Pi^F_{\mathrm{e}}
		&=
		\cosh F \cos(\rho/2) - 1 
		\,,&
		\Pi^{\,\rho}_{\mathrm{e}}
		&=
		2\sinh F \sin(\rho/2)\,,
		\\
		\label{eq:app:Pi_F_o}	
		\Pi^F_{\mathrm{o}}
		&=
		\sinh F \cos(\rho/2) - F
		\,,&
		\Pi^{\,\rho}_{\mathrm{o}}
		&=
		2\cosh F\sin(\rho/2)-\rho\,,
	\end{align}
\end{subequations}
as well as the corresponding retarded and advanced loops \eq{PieoRA},
\begin{subequations}
\begin{align}
	\label{eq:app:PieoR}
  	\Pi_{\mathrm{e/o}}^R(x,y) 
  	& =  \theta(x_{0}-y_{0})\, \Pi_{\mathrm{e/o}}^{\,\rho}(x,y)
	\,,
  	\\
	\label{eq:app:PieoA}
  	\Pi_{\mathrm{e/o}}^A(x,y) 
  	& =  -\theta(y_{0}-x_{0})\, \Pi_{\mathrm{e/o}}^{\,\rho}(x,y)
	 \,.
\end{align}
\end{subequations}
These loop functions are an essential ingredient to the non-perturbative integrals $I$,
\eq{IntegralsIJH}, which can also be decomposed into statistical and spectral components. 
For this, we re-write the integrals in terms of implicit integral equations and after decomposing these, re-express them in terms of non-perturbative coupling functions. 
Using that the coupling function \eq{Lambdae} can be written in two different ways,
\begin{align}
	\Lambda_{\mathrm{e}}(x,y)
	&= \left[{1-\mathrm{i}\bar\lambda\, \Pi_{\mathrm{e}}}\right]^{-1}\!\!(x,y)\,\bar\lambda(y)
	= \bar\lambda(x)\,\left[{1-\mathrm{i}\Pi_{\mathrm{e}}\,\bar\lambda}\right]^{-1}\!\!(x,y)
	\,,
\end{align}
and, in the same way, the functions $\Lambda_{\mathrm{o}}$ and $\ols\Lambda_{\mathrm{e}}$, 
we can write the loop chain integrals \eq{IntegralsIJH} as
\begin{align}
\label{eq:IntegralsIJHimplicit}
	&I_{\mathrm{e}}(x,y)
	=
	-\bar\lambda(x)\,\Pi_{\mathrm{e}}(x,y)\,
	\bar\lambda(y)
	\nonumber
	+\mathrm{i}
	\int_{z,\mathcal{C}} 
	I_{\mathrm{e}}(x,z)\,\Pi_{\mathrm{e}}(z,y)\,\bar\lambda(y)
	\,,\\ 
	&I_{\mathrm{o}}(x,y)
	=
	-\bar\lambda(x)\,\Pi_{\mathrm{o}}(x,y)\,
	\bar\lambda(y)
	-\mathrm{i}
	\int_{z,\mathcal{C}} 
	I_{\mathrm{o}}(x,z)\,\Pi_{\mathrm{o}}(z,y)\,\bar\lambda(y)
	\,.
\end{align}
Note that any matrix product on the CTP starting at some initial time $t_{0}$,
$A(x,y) = \int_{z,\mathcal{C}} B(x,z)\,C(z,y)$,
decomposes into
$A^{F}(x,y) 
= -\mathrm{i}\left(
B^{R}\ast C^{F} + B^{F}\ast C^{A}
\right)(x,y)$,
$A^{\,\rho}(x,y) 
= -\mathrm{i}\left(B^{R}\ast C^{R}-B^{A}\ast C^{A}\right)(x,y)$.
Finite integration limits in time are thereby taken into account by the theta function $\theta(x,y)\equiv\theta(x^{0}-y^{0})$, with $\theta^{-}(x)\equiv\theta(-x)$, which defines the retarded and advanced functions
  	$B^R(x,y) 
	=  (\theta\cdot B^{\,\rho})(x,y)$,
  	$B^A(x,y) 
  	 =   -(\theta^-\cdot B^{\,\rho})(x,y)$,
implying that 
  	$B^{\,\rho} (x,y)
	 =  B^R(x,y)-B^A(x,y)$.
In the decomposition of $A^{\,\rho}(x,y)$, 
depending on whether $x_{0}>y_{0}$ or $x_{0}< y_{0}$, either the  convolution of retarded functions or that of advanced functions is non-zero, such that one gets
	$A^{R}(x,y) 
	= -\mathrm{i}\left(B^{R}\ast C^{R}\right)(x,y)$,
	$A^{A}(x,y) 
	= -\mathrm{i}\left(B^{A}\ast C^{A}\right)(x,y)$.

Using these decompositions and the short-hand notation for matrix products in space-time by a star $\ast\,$, cf.~\App{notation}, we can rewrite the loop chain integrals \eq{IntegralsIJHimplicit} as
\begin{subequations}
\begin{align}
	\label{eq:IeoFimpl}
	I^{F}_{\mathrm{e/o}}
	&=
	- \bar\lambda\ast\Pi^{F}_{\mathrm{e/o}}\ast\bar\lambda
	\pm
	\left(
	  I^{R}_{\mathrm{e/o}}\ast\Pi^{F}_\mathrm{e/o}
	+I^{F}_{\mathrm{e/o}}\ast\Pi^{A}_\mathrm{e/o}
	\right)\ast\bar\lambda
	\,,\\ 
	\label{eq:Ieorhoimpl}
	I^{\,\rho}_{\mathrm{e/o}}
	&=
	- \bar\lambda\ast\Pi^{\,\rho}_{\mathrm{e/o}}\ast\bar\lambda
	\pm 
	\left(
	  I^{R}_{\mathrm{e/o}}\ast\Pi^{R}_\mathrm{e/o}
	-I^{A}_{\mathrm{e/o}}\ast\Pi^{A}_\mathrm{e/o}
	\right)\ast\bar\lambda
	\,,\\ 
	\label{eq:IeoRAimpl}
	I^{R,A}_{\mathrm{e/o}}
	&=
	- \left(\bar\lambda\ast
	\Pi^{R,A}_{\mathrm{e/o}}
	\mp
	  I^{R,A}_{\mathrm{e/o}}\ast\Pi^{R,A}_\mathrm{e/o}
	\right)\ast\bar\lambda
\end{align}
\end{subequations}
where, for keeping the notation compact, it is implied that 
$\rho_{\mathrm{e}}=G^{R,A}_{\mathrm{e}}=0$, $\rho_{\mathrm{o}}=\rho$, $G^{R,A}_{\mathrm{o}}=G^{R,A}$, $F_{\mathrm{e}}=1$, $F_{\mathrm{o}}\equiv F$, and that $\bar\lambda\equiv\bar\lambda(x)\,\delta(x-y)$ is a diagonal matrix.
We note that the dressed coupling $\bar\lambda$, cf.~Eqs.~\eq{lambdabar}, \eq{lambdabarf}, depends only on the local statistical function $G(x,x)\equiv F(x,x)$ and thus is not further affected by the decomposition, cf.~\Eq{lambdabarf}.

We can now write the loop chain integrals again in explicitly re-summed form.
Solving \Eq{IeoRAimpl} with respect to $I^{R,A}_{\mathrm{e/o}}$ yields
\begin{align}
	\label{eq:IeoRAexpl}
	I^{R,A}_{\mathrm{e/o}}
	&=
	- \bar\lambda\ast
	\Pi^{R,A}_{\mathrm{e/o}}
	\ast\bar\lambda
	\ast\left(
	1\mp \Pi^{R,A}_\mathrm{e/o}
	\bar\lambda
	\right)^{-1}
	\,,
\end{align} 
such that, applying the identity $I^{\,\rho}=  I^R-I^A$, one arrives at 
\begin{align}
	\label{eq:Ieorhoexpl}
	I^{\,\rho}_{\mathrm{e/o}}
	&=
	- \bar\lambda\ast
	\left(1\mp \Pi^{R}_\mathrm{e/o}\bar\lambda\right)^{-1}\!\!
	\ast\Pi^{\,\rho}_{\mathrm{e/o}}
	\ast\left(1\mp \bar\lambda\,\Pi^{A}_\mathrm{e/o}\right)^{-1}
	\ast\bar\lambda
	\,.
\end{align} 
Inserting \eq{IeoRAexpl} into \Eq{IeoFimpl} and solving with respect to $I^{F}_{\mathrm{e/o}}$, one finds
\begin{align}
	\label{eq:IeoFexpl}
	&I^{F}_{\mathrm{e/o}}
	=
	- \bar\lambda\ast
	\left(1\mp \Pi^{R}_\mathrm{e/o}\bar\lambda\right)^{-1}\ast
	\Pi^{F}_{\mathrm{e/o}}
	\ast\left(1\mp \bar\lambda\,\Pi^{A}_\mathrm{e/o}\right)^{-1}
	\ast\bar\lambda
	\,.
\end{align} 
In a more compact notation the above integrals and non-perturbative couplings then take the form given in Eqs.~\eq{IntegralsIJHFrho} and \eq{LambdaeoRA}, respectively.


\end{appendix}



%

\end{document}